\documentclass[aps,pra,reprint,superscriptaddress,longbibliography]{revtex4-1}
\usepackage{graphicx,color}
\usepackage{amsmath, amssymb}
\usepackage{longtable}
\usepackage{natbib}
\usepackage{bm}% bold math

\usepackage{hyperref}% add hypertext capabilities
\allowdisplaybreaks[1]

\begin{document}
\title{${\tilde{J}}$-pseudospin states and the crystal field of cubic systems}
\author{Naoya Iwahara}
\email{naoya.iwahara@gmail.com}
\affiliation{Theory of Nanomaterials Group, University of Leuven, Celestijnenlaan 200F, B-3001 Leuven, Belgium}
\author{Liviu Ungur}
%\email{chmlu@nus.edu.sg}
\affiliation{Theory of Nanomaterials Group, University of Leuven, Celestijnenlaan 200F, B-3001 Leuven, Belgium}
\affiliation{Department of Chemistry, National University of Singapore, Block S8 Level 3, 3 Science Drive 3, Singapore 117543}
\author{Liviu F. Chibotaru}
\email{liviu.chibotaru@gmail.com}
\affiliation{Theory of Nanomaterials Group, University of Leuven, Celestijnenlaan 200F, B-3001 Leuven, Belgium}
\date{\today}

\begin{abstract}
Theory of $\tilde{J}$-pseudospin for $f$ element in cubic environment is developed.
By fulfilling the symmetry requirements and the adiabatic connection to atomic limit, the crystal-field states are uniquely transformed into $\tilde{J}$-pseudospin states.
In terms of the pseudospin operators, both the total angular momentum and the crystal-field Hamiltonian contain higher-rank tensor terms than the traditional ones do, which means the present framework naturally include the effects such as the covalency and $J$-mixing beyond the $f$-shell model.
Combining the developed theory with {\it ab initio} calculations, the $\tilde{J}$-pseudospin states for Nd$^{3+}$ and Np$^{4+}$ ions in octahedral sites of insulators are derived. 
\end{abstract}

\maketitle

\section{Introduction}
\label{Sec:Intro}
Crystal-field theory \cite{Bethe1929} has been widely used for the  investigation of the electronic, magnetic, and  optical properties of metal ions in complexes and solids \cite{Abragam1970, Dieke1967}, and it is still intensively used \cite{Ishikawa2003, Magnani2005, McNulty2016, Ruminy2016}.
Although the traditional electrostatic approach seems to provide basic character of the electronic structures, 
as is well known, it does not take account of various effects such as covalency \cite{VanVleck1935, Jorgensen1963}, $J$-mixing \cite{Dieke1967}, and shielding \cite{Watson1967}.
To address accurately the properties of electronic states in metal ions, state-of-the-art {\it ab initio} quantum chemistry methodology including covalency, 
electron correlation, spin-orbit coupling and other relativistic effects is nowadays an alternative popular approach.
Indeed, recently post Hartree-Fock methods are 
starting to be applied to the study of strongly correlated materials containing heavy $d$ elements \cite{Bogdanov2015, Lafrancois2016}.
A common problem of {\it ab initio} approaches is that the computed electronic states do not directly provide a clear physical picture. 
For example, in the case of magnetic systems, they are characterized in terms of pseudospin Hamiltonian \cite{Abragam1970}.
While the {\it ab initio} states must contain all necessary information, it is not {\it a priori} clear how to extract the pseudospin Hamiltonian on their basis. 

This issue has been recently addressed, and general principles
for the derivation of the uniquely defined pseudospin 
Hamiltonian
from {\it ab initio} calculated electronic states was proposed \cite{Chibotaru2008, Chibotaru2012, Chibotaru2013}:
the principles consist of (1) symmetry requirements and (2) adiabatic connection to the well-defined limiting cases. 
$N$
low-energy electronic states is selected for the description of low-energy phenomena, 
the $\tilde{S}$-pseudospin states ($N = 2\tilde{S}+1$) are derived by an unitary transformation of these electronic states
and then, the pseudospin Hamiltonian is derived using the obtained pseudospin states. 
The unitary matrix should be uniquely determined based on these principles. 
There is no difficulty for the unique definition of small pseudospins ($\tilde{S} = 1/2$ and 1): 
Indeed, when only the small pseudospins are relevant,
combining the 
theoretical framework with {\it ab initio} calculations, various magnetic properties of metal complexes have been explained \cite{Chibotaru2008angew, Chibotaru2015} and predicted \cite{Ungur2014, Ungur2016Strategy}.
On the other hand, the derivation of large pseudospin $\tilde{S} \ge 3/2$, which is relevant to e.g. $\tilde{J}$-pseudospin for the crystal-field states of $f$-elements, remains under development \cite{Ungur2017, Gamma8} because a practical algorithm to determine a large number of the unitary matrix elements ($\approx N^2/2$) fulfilling both requirements is not obvious. 

In this work, we develop the methodology to uniquely transform the crystal-field states of $f$ elements in cubic environment into the $\tilde{J}$-pseudospin states satisfying the symmetry 
requirements
and the adiabatic connection between the $\tilde{J}$-pseudospin states and the corresponding atomic $J$-multiplet.
The present $\tilde{J}$-pseudospin states naturally include the effects beyond the traditional crystal-field model based on isolated $f$ orbitals, 
resulting in the presence of the higher rank tensor terms in total angular momentum and crystal-field Hamiltonian than in conventional approaches based on atomic 
$J$-multiplet. 
The developed theory is applied to Nd$^{3+}$ and Np$^{4+}$ ions in cubic environment.

\section{Unique definition of pseudospin}
\label{Sec:pseudospin}
For the description of the local electronic structure and properties of magnetic ions, phenomenological pseudospin Hamiltonians are often employed \cite{Abragam1970}
\footnote{
Since pure spin/orbital/total angular momentum operators do not commute with the Hamiltonian of materials due to the coexistence of crystal-field and spin-orbit coupling, the ``spin'' operators in phenomenological model are not pure ones but correspond to pseudo (effective, fictious) spin (see Sec. 1.4 and 3.1 in Ref. \cite{Abragam1970}).
}. 
The pseudospin Hamiltonian acts on the abstract pseudospin states $|\tilde{S}M\rangle$ ($M = -\tilde{S}, -\tilde{S}+1, \cdots, \tilde{S}$), and its eigenstates describe the low-energy states. 
On the other hand, if the exact electronic states responsible for the low-energy phenomena of interest are given, 
\begin{eqnarray}
\mathcal{H} = \{|\Psi_i\rangle| i = 1, 2, \cdots, N\},
\label{Eq:H}
\end{eqnarray}
the pseudospin states $|\tilde{S}M\rangle$ should be obtainable directly from this set of states.
However, the relation between them is not {\it a priori} evident. 
This problem has been recently addressed by some of us and the methodology to uniquely define the pseudospin states was proposed \cite{Chibotaru2008, Chibotaru2012, Chibotaru2013}.

The pseudospin states may be obtained by unitary transformation of the electronic 
states
$|\Psi_i\rangle$:
\begin{eqnarray}
 |\tilde{S}M\rangle &=& \sum_{i=1}^N U_{iM} |\Psi_i\rangle,
\label{Eq:SM}
\end{eqnarray}
where, $U_{iM}$ are elements of a unitary matrix $U$ and $N = 2\tilde{S}+1$. 
Once pseudospin states are established, the pseudospin operators such as 
\begin{eqnarray*}
\tilde{S}_z = \sum_{M=-\tilde{S}}^{\tilde{S}} M|\tilde{S}M\rangle \langle \tilde{S}M|, 
\end{eqnarray*}
and irreducible tensor operators $Y_{kq}(\tilde{\bm{S}})$ can be assigned in their basis, 
where, $k$ and $q$ indicate the rank and the component of the tensor, respectively.
Nevertheless, for an arbitrary choice of $U$, the obtained operators $\tilde{\bm{S}}$ would not behave as expected for the phenomenological effective spin 
under symmetry operations, and the obtained pseudospin Hamiltonian will also differ from the phenomenological one.
In order to choose adequate unitary transformation $U$ in Eq. (\ref{Eq:SM}), two requirements (principles) are employed \cite{Chibotaru2008, Chibotaru2012, Chibotaru2013}:
\begin{enumerate}
 \item The pseudospin states $|\tilde{S}M\rangle$ transform as the true spin states $|SM\rangle$ ($S = \tilde{S}$) under the time-reversal and spatial symmetry operations. 
 \item The pseudospin states are adiabatically connected to the well-defined pure spin/orbital/total angular momentum states. 
\end{enumerate}
The first principle simply requires the pseudospin states to be consistent with the 
symmetries
of the system \cite{Abragam1970}
\footnote{
In order to fulfill time-reversal symmetry, it is necessary to generate integer and half-integer pseudospin states for non-Kramers and Kramers systems, respectively.
However, sometimes half-integer pseudospin is used to describe (quasi) degenerate state of non-Kramers system for simple description of the Hamiltonian.
Care is needed in this case because not all symmetry properties are fulfilled, therefore, its unusual behaviour might be expected.
For example, double degenerate $\Gamma_3$ states never transform as $\Gamma_6$ ($S=1/2$ spin) states in cubic group.
Even if the pseudospin states are enforced to fulfill the time-reversal symmetry, the crystal-field Hamiltonian contains odd order terms of pseudospin operator \cite{Mueller1968}.
}.
The second principle requires the existence of the one-to-one correspondence between the pseudospin and a well-defined pure spin.
This correspondence is established by adiabatically turning on the interaction which only exists in the materials \cite{Chibotaru2008, Chibotaru2013}. 
The latter may include covalency, spin-orbit coupling and deformation of the environment, depending on the choice of the reference situation. 
Such an adiabatic connection is used in various fields of condensed matter physics to characterize the systems \cite{Anderson, Laughlin}.

The proposed principles state the requirements for the unique definition of pseudospins, while they do not provide the practical way to achieve it.
In practice, low-dimensional pseudospins ($\tilde{S}=$ 1/2, 1) can be uniquely defined by identifying their states with the Zeeman states along one of the principal magnetic axes of the system \cite{Chibotaru2012, Chibotaru2013}.
These pseudospin states obey automatically the symmetry requirements of principle 1. 
On the other hand, the unique definition of larger pseudospin $\tilde{S} \ge 3/2$ is technically more difficult than that of small pseudospins due to the quadratically increasing number of free parameters ($\propto N^2$) defining the unitary transformation $U$ in Eq. (\ref{Eq:SM}) \cite{Chibotaru2013}.
If as in the small pseudospins, the eigenstates of 
the magnetic moment
$\hat{\mu}_Z$ along principal magnetic axis $Z$
are taken as pseudospin states \cite{Ungur2017}, the spatial symmetry requirement may not be completely fulfilled.  
For example, the crystal field states of a Kramers ion in cubic environment may contain four-fold degenerate $\Gamma_8$ states (Table \ref{Table:J_Gamma_Bk}), whereas the eigenstates of 
$\hat{\mu}_Z$ 
never do so because they satisfy at most a tetragonal symmetry under Zeeman splitting. 
Although the definition 
of the pseudospins
via eigenstates of 
$\hat{\mu}_Z$ 
is one of the possible choices,
the obtained Hamiltonian will not have {\it a priori} the expected form for cubic system.
Another issue is the requirement of the adiabatic connection:
this can be in principle satisfied by defining the pseudospin by several consecutive {\it ab initio} calculations in which some controlling parameters are varied (see Ref. \cite{Chibotaru2008} and Sec. VI in Ref. \cite{Chibotaru2013}). 
It is evident that such brute force approach is far from practical for most of systems of interest. 
Towards the establishment of the practical scheme to determine large pseudospins, the theory of the $\tilde{J}$-pseudospin in cubic environment is developed below.

\begin{table*}[tb]
\begin{ruledtabular}
\caption{
The relation between $J$, its decomposition into $\Gamma$ irreducible representations of cubic group $\mathcal{G}$ ($= O, O_h, T_d$), 
and crystal-field parameters $B_k$ in cubic environment. 
$f$-ions (Ln: lanthanide, Ac: actinide \cite{Edelstein2006}) whose ground atomic multiplets are characterized by $J$ are also shown.
Parity ($g$ or $u$) is not shown.
}
\label{Table:J_Gamma_Bk}
\begin{tabular}{cccccc}
$J$ & $f^n$ & Ln & Ac & $J\downarrow \mathcal{G}$ & $B_k$ \\
\hline
0    & & & & $\Gamma_1$ & - \\         
1    & & & & $\Gamma_4$ & - \\         
2    &                 &                                 &  
 & $\Gamma_3 \oplus \Gamma_5$                                                  & $B_4$ \\
3    &                 &                                 &  
 & $\Gamma_2 \oplus \Gamma_4 \oplus \Gamma_5$                                  & $B_4, B_6$ \\
4    & $f^2$, $f^4$    & Pr$^{3+}$, Pm$^{3+}$            & U$^{4+}$, Np$^{3/5+}$, Pu$^{4/6+}$ 
 & $\Gamma_1 \oplus \Gamma_3 \oplus \Gamma_4 \oplus \Gamma_5$                  & $B_4, B_6, B_8$ \\
5    &                 &                                 &  
 & $\Gamma_3 \oplus 2\Gamma_4 \oplus \Gamma_5$                                 & $B_4, B_6, B_8, B_{10}$ \\
6    & $f^8$, $f^{12}$ & Tb$^{3+}$, Tm$^{3+}$            & Bk$^{3+}$, Cf$^{4+}$
 & $\Gamma_1 \oplus \Gamma_2 \oplus \Gamma_3 \oplus \Gamma_4 \oplus 2\Gamma_5$ & $B_4, B_6, B_8, B_{10}, B_{12}$ \\
7    &                 &                                 &  
 & $\Gamma_2 \oplus \Gamma_3 \oplus 2\Gamma_4 \oplus 2\Gamma_5$                & $B_4, B_6, B_8, B_{10}, B_{12}, B_{14}$ \\ 
8    & $f^{10}$        & Ho$^{3+}$                       & Es$^{3+}$ 
 & $\Gamma_1 \oplus 2\Gamma_3 \oplus 2\Gamma_4 \oplus 2\Gamma_5$               & $B_4, B_6, B_8, B_{10}, B_{12}, B_{14}, B_{16}$ \\
1/2  & & & & $\Gamma_6$ & - \\         
3/2  & & & & $\Gamma_8$ & - \\         
5/2  & $f^1$, $f^5$    & Ce$^{3+}$, Sm$^{3+}$, Pr$^{4+}$ & Pa$^{4+}$, U$^{5+}$, Pu$^{3+}$, Am$^{4+}$ %Th$^{3+}(5d^1)$, 
 & $\Gamma_7 \oplus \Gamma_8$                                                  & $B_4$ \\
7/2  & $f^{13}$        & Yb$^{3+}$                       &  
 & $\Gamma_6 \oplus \Gamma_7 \oplus \Gamma_8$                                  & $B_4, B_6$ \\
9/2  & $f^3$           & Nd$^{3+}$                       & U$^{3+}$, Np$^{4+}$, Pu$^{5+}$
 & $\Gamma_6 \oplus 2\Gamma_8$                                                 & $B_4, B_6, B_8$ \\
11/2 &                 &                                 &  
 & $\Gamma_6 \oplus \Gamma_7 \oplus 2\Gamma_8$                                 & $B_4, B_6, B_8, B_{10}$ \\
13/2 &                 &                                 &  
 & $\Gamma_6 \oplus 2\Gamma_7 \oplus 2\Gamma_8$                                & $B_4, B_6, B_8, B_{10}, B_{12}$ \\
15/2 & $f^9$, $f^{11}$ & Dy$^{3+}$, Er$^{3+}$            & Cf$^{3+}$, Es$^{2+}$
 & $\Gamma_6 \oplus \Gamma_7 \oplus 3\Gamma_8$                                 & $B_4, B_6, B_8, B_{10}, B_{12}, B_{14}$ \\
\end{tabular}
\end{ruledtabular}
\end{table*}

\section{Pseudospin in cubic environment}
\label{Sec:pseudospin_cubic}
The low-energy crystal-field states of $f$ elements mainly originate from the ground atomic $J$-multiplet \cite{Abragam1970}.
Thus, the crystal-field Hamiltonian is described in terms of $\tilde{J}$-pseudospin operators. 
Here, the algorithm to derive the $\tilde{J}$-pseudospin crystal-field Hamiltonian in octahedral environment from the crystal-field states is shown taking $\tilde{J}=9/2$ pseudospin as an example because the latter is the simplest non-trivial case
where both requirements in Sec. \ref{Sec:pseudospin} have to be fully taken into account. 
Other cases can be done using the 
formulae
in Appendix \ref{A:JM}.
The developed method is applied to derive the crystal-field Hamiltonian of Nd$^{3+}$ ($4f^3$) and Np$^{4+}$ ($5f^3$) ions in octahedral 
environment.

\subsection{$\Gamma$-pseudospin}
In an octahedral ($O$ or $O_h$) environment, %the low-lying crystal-field levels which mainly come from 
the ground atomic $J=9/2$ multiplets split into two sets of four-fold degenerate $\Gamma_8$ multiplets and one $\Gamma_6$ Kramers doublet (Table \ref{Table:J_Gamma_Bk}).
Since the $\Gamma_8$ and $\Gamma_6$ states, respectively, transform as $S=3/2$ and $S=1/2$ spin states under the symmetry operations of the $O_h$ group \cite{Koster1963},
each of the multiplets can be unambiguously transformed into $\Gamma$-pseudospin state by requirement 1 
[$\mathcal{H}$ corresponds to a set of degenerate $\Gamma$ states]
\cite{Chibotaru2013, Gamma8}.
Hereafter, the three $C_4$ axes of the cubic environment correspond to the $x, y, z$ axes (right-handed coordinate system), the $z$ axis is taken as the quantization axis of the angular momentum, and the basis of the irreducible representations given in Ref. \cite{Koster1963} is used.
Using the generators of the rotational symmetry operations of the $O_h$ group, for example, $\pi/2$ rotations around the $y$ and $z$ axes ($\hat{C}_4^y$ and $\hat{C}_4^z$), the $\Gamma$ multiplets are transformed as, respectively, 
\begin{eqnarray}
 \hat{C}_4^y |\Gamma M\rangle &=& \sum_{M'} d^{\tilde{S}}_{M'M}\left(\frac{\pi}{2}\right)|\Gamma M'\rangle,
\label{Eq:C4y}
\end{eqnarray}
and 
\begin{eqnarray}
 \hat{C}_4^z |\Gamma M\rangle &=& e^{-i\frac{\pi}{2}M}|\Gamma M\rangle.
\label{Eq:C4z}
\end{eqnarray}
Here, $\tilde{S} = 1/2$ for $\Gamma = \Gamma_6$, $\tilde{S} = 3/2$ for $\Gamma = \Gamma_8$, $M, M' = -\tilde{S}, -\tilde{S} + 1, ..., \tilde{S}$,
and $d^{\tilde{S}}_{M'M}$ is the rotation matrix around the $y$ axis (Wigner $D$-function) \cite{Varshalovich1988}.
The relative phase factors between $|\Gamma M\rangle$'s are fixed by using time-reversal symmetry \cite{Abragam1970, Koster1963, Huby1954}:
\begin{eqnarray}
 \hat{\theta} |\Gamma M\rangle &=& (-1)^{\tilde{S} - M} |\Gamma, -M\rangle,
\label{Eq:time}
\end{eqnarray}
where $\hat{\theta}$ is the time-reversal operator.
Similar consideration holds for a $T_d$ system by replacing $C_4$ with $S_4$.

\subsection{$\tilde{J}$-pseudospin}
\label{Sec:Jpseudospin}
The $\tilde{J}$-pseudospin states are described by linear combinations of the $\Gamma$-pseudospin states
[$\mathcal{H} = \{|\Gamma M\rangle| \Gamma = \Gamma_6, \Gamma_8^{(1)}, \Gamma_8^{(2)}\}$]:
\begin{eqnarray}
 |\tilde{J} M\rangle &=& \sum_{\mu \Gamma M'} U_{\Gamma^{(\mu)} M', \tilde{J}M} |\Gamma^{(\mu)} M'\rangle,
\label{Eq:JM}
\end{eqnarray}
where, the index $\mu$ distinguishes the repeated $\Gamma$ multiplets (two $\Gamma_8$ states in the present case), and $U_{\Gamma^{(\mu)} M, \tilde{J}M}$ are coefficients. 
The latter are restricted by the first requirement.
$|\tilde{J}M\rangle$ with $M = \mp 7/2, \pm 1/2, \pm 9/2$ transform as $|\Gamma, \pm 1/2\rangle$  under the $C_4^z$ rotation. 
The relation between the $|\tilde{J}, M\rangle$ states and $|\Gamma_6, \pm 1/2\rangle$ states is unambiguously given by taking account of the transformations under $C_4^y$ rotation. 
On the other hand, the relation between the $|\tilde{J}M\rangle$ and two $|\Gamma_8 \pm 1/2\rangle$ states is given up to the arbitrary mixing (rotation) of the two $\Gamma_8$ states described by one angle $\alpha$.
Finally, making use of the components of $|\Gamma_8, \pm 3/2\rangle$ appearing in $\hat{C}_4^y|\Gamma_8, \pm 1/2\rangle$, the unitary matrix $U$ in Eq. (\ref{Eq:JM}) is determined up to angle $\alpha$. 
The obtained $\tilde{J}=9/2$ pseudospin states are
\begin{widetext}
\begin{eqnarray}
 \left|\tilde{J}, \mp \frac{9}{2} (\alpha) \right\rangle &=& 
 \frac{1}{2}\sqrt{\frac{3}{2}}  \left|\Gamma_6, \mp \frac{1}{2}\right\rangle
 \mp  \frac{1}{2} \sqrt{\frac{5}{2}}
  \left[
    \cos \alpha \left|\Gamma_8^{(1)}, \mp \frac{1}{2}\right\rangle 
  - \sin \alpha \left|\Gamma_8^{(2)}, \mp \frac{1}{2}\right\rangle
  \right],
\nonumber
\\
 \left|\tilde{J}, \mp \frac{7}{2} (\alpha) \right\rangle &=& 
  \frac{1}{2\sqrt{6}} \left|\Gamma_6, \pm \frac{1}{2}\right\rangle
  \pm \frac{1}{2} \sqrt{\frac{23}{6}}
  \left[
   \sin(\alpha + \phi_1) \left|\Gamma_8^{(1)}, \pm \frac{1}{2}\right\rangle 
 - \cos(\alpha + \phi_1) \left|\Gamma_8^{(2)}, \pm \frac{1}{2}\right\rangle
  \right],
\nonumber
\\ 
 \left|\tilde{J}, \mp \frac{5}{2} (\alpha) \right\rangle &=& 
 \pm \left[
    \sin(\alpha + \phi_2) \left|\Gamma_8^{(1)}, \pm \frac{3}{2}\right\rangle 
  + \cos(\alpha + \phi_2) \left|\Gamma_8^{(2)}, \pm \frac{3}{2}\right\rangle 
 \right],
\nonumber
\\
 \left|\tilde{J}, \mp \frac{3}{2} (\alpha) \right\rangle &=& 
 \pm \left[
   -\cos(\alpha + \phi_2) \left|\Gamma_8^{(1)}, \mp \frac{3}{2}\right\rangle 
  + \sin(\alpha + \phi_2) \left|\Gamma_8^{(2)}, \mp \frac{3}{2}\right\rangle 
 \right],
\nonumber
\\
 \left|\tilde{J}, \mp \frac{1}{2} (\alpha) \right\rangle &=& 
  \frac{1}{2} \sqrt{\frac{7}{3}} \left|\Gamma_6, \mp \frac{1}{2}\right\rangle
  \pm \frac{1}{2}\sqrt{\frac{5}{3}} \left[
    \sin(\alpha + \phi_3) \left|\Gamma_8^{(1)}, \mp \frac{1}{2}\right\rangle
   +\cos(\alpha + \phi_3) \left|\Gamma_8^{(2)}, \mp \frac{1}{2}\right\rangle
  \right],
\label{Eq:pseudoJ92}
\end{eqnarray}
\end{widetext}
where, % $\alpha$ is an arbitrary real parameter, %$\alpha = \arcsin(\sqrt{7/10})$, and $\beta = \arcsin(5/\sqrt{46})$.
$\phi_1 = \arccos \sqrt{3/115}$, 
$\phi_2 = \arccos \sqrt{7/10}$, 
and $\phi_3 = \arccos (2/5)$.
%As in Eq. (\ref{Eq:time}) for the $\Gamma$-pseudospin states, the phase factors of $\tilde{J}$-pseudospin states are determined to satisfy
The phase factors of $\tilde{J}$-pseudospin states are determined to satisfy $\hat{\theta} |\tilde{J}M\rangle = (-1)^{\tilde{J}-M}|\tilde{J},-M\rangle$ under time-inversion as in Eq. (\ref{Eq:time}) for $\Gamma$-pseudospin states (see for the phase factors and time-reversal symmetry Ref. \cite{Huby1954}).
The angle $\alpha$ is explicitly present in the left hand sides of Eq. (\ref{Eq:pseudoJ92}) because it is not fixed yet. 
In addition to $\alpha$, there are two possibilities for the assignment of two $\Gamma_8$ states in $\mathcal{H}$.
By the similar procedures, all the important cases for $f$ elements can be derived (see Appendix \ref{A:JM}).

Using the pseudospin states (\ref{Eq:pseudoJ92}), we can define the irreducible tensor operators (Appendix \ref{A:ITO})
\begin{eqnarray}
 \mathcal{Y}_{kq}(\tilde{\bm{J}}(\alpha)) 
 &=&
 \frac{Y_{kq}(\tilde{\bm{J}}(\alpha))}{Y_{k0}(\tilde{J})} 
\nonumber\\
% &=& \sum_{M,M'=-\tilde{J}}^{\tilde{J}}
 &=& \sum_{M,M'}
%     \frac{C_{\tilde{J}M kq}^{\tilde{J}M'}}{C_{\tilde{J}\tilde{J}k0}^{\tilde{J}\tilde{J}}} 
     \frac{\langle (\tilde{J}k) \tilde{J}M'| \tilde{J}M kq \rangle}{\langle (\tilde{J}k) \tilde{J}\tilde{J}| \tilde{J}\tilde{J} k0 \rangle}
     |\tilde{J} M' (\alpha)\rangle \langle \tilde{J} M (\alpha)|.
\nonumber\\
\label{Eq:Ykq}
\end{eqnarray}
Here, $\tilde{\bm{J}}$ is the $\tilde{J}$-pseudospin operator,  
%$C_{a\alpha b\beta}^{c\gamma} = \langle c\gamma|a\alpha b\beta \rangle$ are Clebsch-Gordan coefficients, 
$Y_{kq}(\tilde{\bm{J}})$ is the irreducible tensor operator of rank $k$ ($k = 0, 1, ..., 2\tilde{J}$) and argument $q$ ($q = -k, -k+1, ..., k$),
$Y_{k0}(\tilde{J}) = \langle \tilde{J}\tilde{J}|Y_{k0}(\tilde{\bm{J}})|\tilde{J}\tilde{J}\rangle$,
and $\langle (j_1 j_2) jm| j_1 m_1 j_2 m_2\rangle$ are Clebsch-Gordan coefficients \cite{Varshalovich1988}. 
%Eq. (\ref{Eq:Ykq}) includes the $\tilde{J}$-pseudospin operator as the first rank $\tilde{J}_q = \tilde{J} \mathcal{Y}_{1q}(\tilde{\bm{J}})$.
The tensor operator behaves as a pseudospin state $|\tilde{J}=k, M=q\rangle$ under time-inversion, $\mathcal{Y}_{kq} \rightarrow (-1)^{k-q}\mathcal{Y}_{k,-q}$.
%With $\mathcal{Y}_{kq}$, %the irreducible tensor operators (\ref{Eq:Ykq}), 
Any electronic operators acting on the crystal-field states in $\mathcal{H}$ can be decomposed into $\mathcal{Y}_{kq}$'s (see Appendix \ref{A:ITO}).

For the unique definition of $\tilde{J}$-pseudospin, the variable $\alpha$ in Eq. (\ref{Eq:pseudoJ92}) has to be fixed.
To this end, the second principle is used.
The $\tilde{J}$-pseudospin states (\ref{Eq:pseudoJ92}) and thus $\tilde{\bm{J}}$ have to converge to the atomic $J$-multiplet and pure total angular momentum $\hat{\bm{J}}$, respectively, by adiabatically reducing the interactions with the environment. 
%Therefore, the angle $\alpha$ has to be chosen so that $\tilde{\bm{J}}$ can converge to $\hat{\bm{J}}$ by reducing the effect of the environment. 
%The total angular momentum operator of the system is decomposed as 
This is achieved by choosing $\alpha$ so that the first rank parameter of $\hat{J}_z$, $j_{10}(\alpha)$, becomes the largest:
\begin{eqnarray}
 \hat{J}_z &=& \sum_{k=1}^{2\tilde{J}} \sum_{q=-k}^k j_{kq}(\alpha) \mathcal{Y}_{kq}(\tilde{\bm{J}}(\alpha)).
\label{Eq:J0}
\end{eqnarray}
In general, $j_{10} < \tilde{J}$ because the degree of the mixing of the atomic $J$-multiplets $|JM\rangle$ to the crystal-field states $|\Psi_i\rangle$ depends on $M$ owing to e.g., the covalency and $J$-mixing. 
Substituting $\alpha_0$ maximizing $j_{10}(\alpha)$ into Eq. (\ref{Eq:pseudoJ92}), the $\tilde{J}$-pseudospin states are uniquely defined. 
In this procedure, all possible assignments of $\Gamma_8$ crystal-field levels to $\Gamma^{(1)}_8$ and $\Gamma^{(2)}_8$ in Eq. (\ref{Eq:JM}) also have to be examined. 
If other angle $\alpha$ such as the one at the other extremum is chosen, $\tilde{\bm{J}}$ does not converge to $\hat{\bm{J}}$ in the atomic limit (see Sec. \ref{Sec:f3}) because such choice makes $|\tilde{J}M(\alpha)\rangle$ dissimilar from $|JM\rangle$.
The same procedure uniquely defines $\tilde{J} \ge 9/2$, whereas the $\tilde{J} < 9/2$ pseudospin states are uniquely defined by symmetry.

With the use of the $\mathcal{Y}_{kq}(\tilde{\bm{J}}(\alpha_0))$, the crystal-field Hamiltonian $\hat{H}_\text{cf} = \sum_{\mu \Gamma M} E_{\Gamma}^{(\mu)} |\Gamma^{(\mu)} M \rangle \langle \Gamma^{(\mu)} M|$ is expressed as (see Appendix \ref{A:Hcf}):
\begin{widetext}
\begin{eqnarray}
 \hat{H}_\text{cf} &=& B_{0} 
               + B_4 \left( 
%                  \frac{Y_4^0(\tilde{\bm{J}}(\alpha))}{Y_4^0(\tilde{J})} + \sum_{q = -4,4} \sqrt{\frac{5}{14}} \frac{Y_4^q(\tilde{\bm{J}}(\alpha))}{Y_4^0(\tilde{J})} 
%                  \mathcal{Y}_{40}(\tilde{\bm{J}}) + \sum_{q = \pm 4} \sqrt{\frac{5}{14}} \mathcal{Y}_{4q}(\tilde{\bm{J}})
                  \mathcal{Y}_{40} + \sum_{q = \pm 4} \sqrt{\frac{5}{14}} \mathcal{Y}_{4q}
                 \right)
%+ B_6 \Bigg( \mathcal{Y}_{60}
%\nonumber\\
% &+&
+
                 B_6 \left( 
%                  \frac{Y_6^0(\tilde{\bm{J}}(\alpha))}{Y_6^0(\tilde{J})} - \sum_{q = -4,4} \sqrt{\frac{7}{2}} \frac{Y_6^q(\tilde{\bm{J}}(\alpha))}{Y_6^0(\tilde{J})} 
%                  \mathcal{Y}_{60}(\tilde{\bm{J}}) - \sum_{q = \pm 4} \sqrt{\frac{7}{2}} \mathcal{Y}_{6q}(\tilde{\bm{J}})
                  \mathcal{Y}_{60} - \sum_{q = \pm 4} \sqrt{\frac{7}{2}} \mathcal{Y}_{6q}
                 \right)
%             + B_8 \Bigg( \mathcal{Y}_{80}(\tilde{\bm{J}})
%             + B_8 \Bigg( \mathcal{Y}_{80}
%\right.
%\nonumber\\
%             &+& B_8 \left( 
             + B_8 \left( 
                  \mathcal{Y}_{80}
%&+&
+
%\left.
%                   \sum_{q = \pm 4} \frac{1}{3}\sqrt{\frac{14}{11}} \mathcal{Y}_{8q}(\tilde{\bm{J}})
                   \sum_{q = \pm 4} \frac{1}{3}\sqrt{\frac{14}{11}} \mathcal{Y}_{8q}
%\right.
%\nonumber\\
%&+&
%\left.
+
%                   \sum_{q = \pm 8} \frac{1}{3}\sqrt{\frac{65}{22}} \mathcal{Y}_{8q}(\tilde{\bm{J}})
                   \sum_{q = \pm 8} \frac{1}{3}\sqrt{\frac{65}{22}} \mathcal{Y}_{8q}
                 \right),
%                  \Bigg),
\nonumber\\
\label{Eq:HCF}
\end{eqnarray}
where, 
$\mathcal{Y}_{kq}(\tilde{\bm{J}}(\alpha_0))$ is replaced by $\mathcal{Y}_{kq}$ for simplicity, 
and $B_k$ are calculated as %$B_0 = \frac{1}{10}[2E_{\Gamma_6} + 4(E_{\Gamma_8}^{(1)} + E_{\Gamma_8}^{(2)})]$ and 
\begin{eqnarray}
 B_0 &=& \frac{1}{10} \left[ 2E_{\Gamma_6} + 4(E_{\Gamma_8}^{(1)} + E_{\Gamma_8}^{(2)}) \right],
\nonumber\\
 B_4 &=& \frac{3}{1430}
            \left[ 
             49(2E_{\Gamma_6} - E_{\Gamma_8}^{(1)} - E_{\Gamma_8}^{(2)})
%\right.
%\nonumber\\
% &+&
%\left.
+
              (133 \cos 2\alpha_0 - 4\sqrt{21} \sin 2\alpha_0) (E_{\Gamma_8}^{(1)} - E_{\Gamma_8}^{(2)}) 
            \right],
\nonumber\\
 B_6 &=& \frac{1}{220}
            \left[
             -4(2E_{\Gamma_6} - E_{\Gamma_8}^{(1)} - E_{\Gamma_8}^{(2)})
%\right.
%\nonumber\\
% &+&
%\left.
+
              (8\cos 2\alpha_0 + \sqrt{21}  \sin 2\alpha_0) (E_{\Gamma_8}^{(1)} - E_{\Gamma_8}^{(2)}) 
            \right],
\nonumber\\
 B_8 &=& \frac{1}{1040}
            \left[
             3(2E_{\Gamma_6} - E_{\Gamma_8}^{(1)} - E_{\Gamma_8}^{(2)})
%\right.
%\nonumber\\
% &+&
%\left.
+
              (-3 \cos 2\alpha_0 + 4\sqrt{21} \sin 2\alpha_0) (E_{\Gamma_8}^{(1)} - E_{\Gamma_8}^{(2)}) 
            \right].
\label{Eq:BCF}
\end{eqnarray}
\end{widetext}
Contrary to the conventional crystal-field Hamiltonian containing only fourth and sixth rank terms \cite{Lea1962}, the present one contains up to eighth rank terms (in general up to rank $k \le 2\tilde{J}$).
The conventional form is recovered by imposing the constraint that all local crystal-field levels arise from the atomic $f$ shell.

The proposed algorithm for the unique definition of $\tilde{J}$-pseudospin states in cubic environment is summarized as follows:
\begin{enumerate}
 \item Express $\tilde{J}$-pseudospin states $|\tilde{J}M\rangle$ using Eq. (\ref{Eq:pseudoJ92}) or the corresponding formulae in Appendix \ref{A:JM}.
 \item Maximize the first rank parameter $j_{10}$ of $\hat{J}_z$ (\ref{Eq:J0}) with respect to the free parameters. 
\end{enumerate}
These two procedures satisfy the principles 1 and 2 (Sec. \ref{Sec:pseudospin}), respectively. 
With the obtained $\tilde{J}$-pseudospin states with the fixed angles, any operators acting on the same Hilbert space $\mathcal{H}$ can be decomposed into the irreducible tensor operators $\mathcal{Y}_{kq}$'s (see Appendix \ref{A:ITO}).
In the next section, this algorithm is applied to two systems.

\subsection{{\it Ab initio} derivation of $\tilde{J}=9/2$ pseudospin states}
\label{Sec:Example}
Combining the developed theory and {\it ab initio} calculations, the $\tilde{J}=9/2$ pseudospin states of Nd$^{3+}$ ion in octahedral site of Cs$_2$NaNdCl$_6$ \cite{Zhou2006} and Np$^{4+}$ impurity ion in octahedral Zr site of Cs$_2$ZrCl$_6$ \cite{Bernstein1979, Edelstein1980} are derived.
It is also shown that the present approach fulfills the requirement 2.

\begin{table}[tb]
\caption{Crystal-field levels of Cs$_2$NaNdCl$_6$ and Cs$_2$ZrCl$_6$:Np$^{4+}$ (cm$^{-1}$)
\footnote{$\Gamma_8$ levels within the method (b) are slightly split: 0.2 cm$^{-1}$ and 0.4 cm$^{-1}$ for the ground and excited $\Gamma_8$ levels, respectively.
In this work, the averaged values of the slightly split $\Gamma_8$ levels were used.}.
(a) and (b) indicate the {\it ab initio} methodology (Sec. \ref{Sec:abinitio}) and ``Exp.'' the experimental data \cite{Zhou2006}.
The ground $\Gamma_8$ energy is set to zero.
}
\label{Table:E}
\begin{ruledtabular}
\begin{tabular}{ccccc}
$\Gamma$  & \multicolumn{4}{c}{Energy}\\
          & \multicolumn{3}{c}{Cs$_2$NaNdCl$_6$} & Cs$_2$ZrCl$_6$:Np$^{4+}$ \\
          & (a) & (b) & Exp. \cite{Zhou2006} & (a) \\
\hline
$\Gamma_6$ &   90.225 &  95.318 &  97 &  506.834 \\ 
$\Gamma_8$ &  267.562 & 315.578 & 335 & 1352.775 \\
\end{tabular}
\end{ruledtabular}
\end{table}

\begin{figure*}[tb]
\begin{center}
\begin{tabular}{lll}
(a) & ~ & (b) \\
\includegraphics[height=4.5cm]{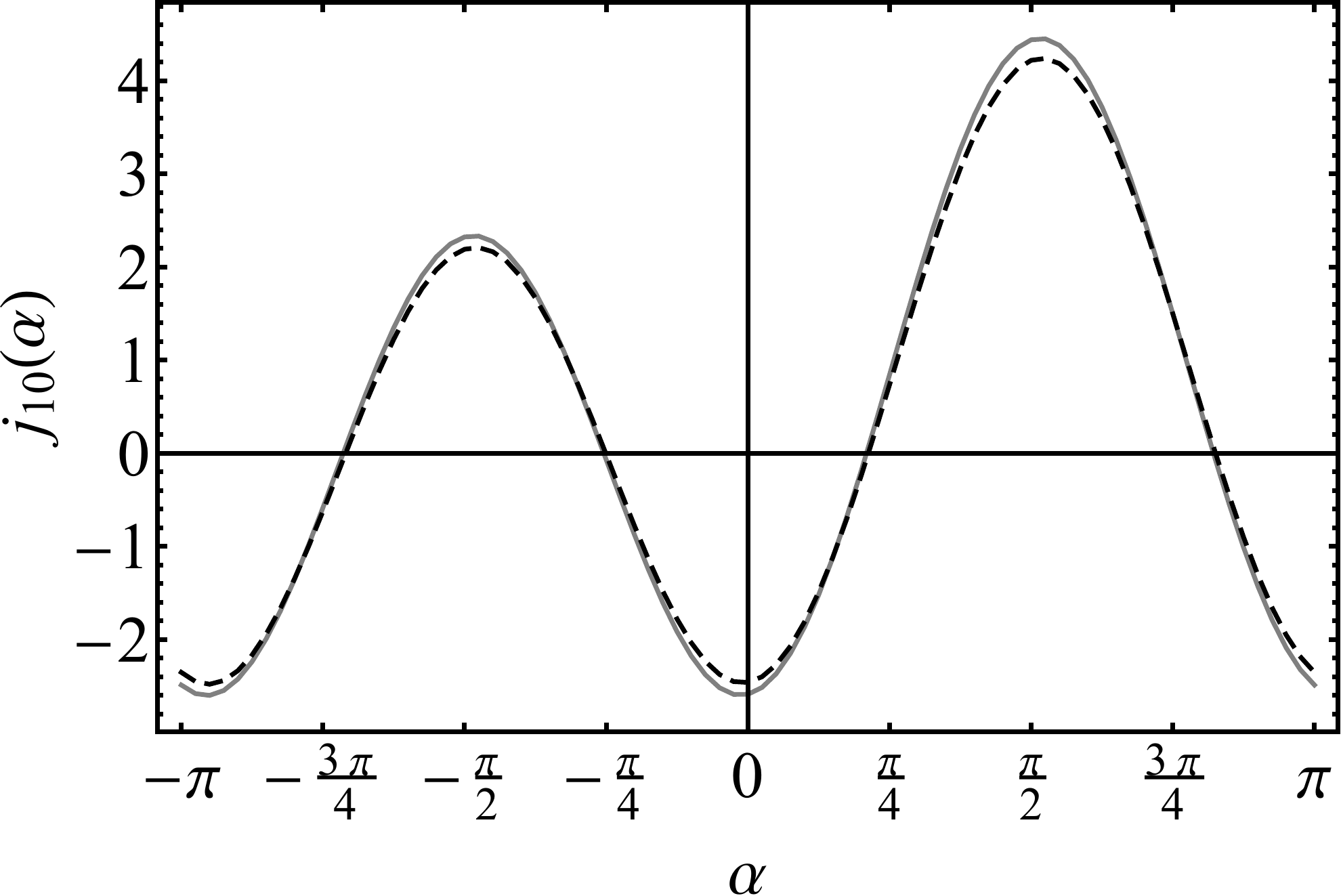}
&
&
\includegraphics[height=4.5cm]{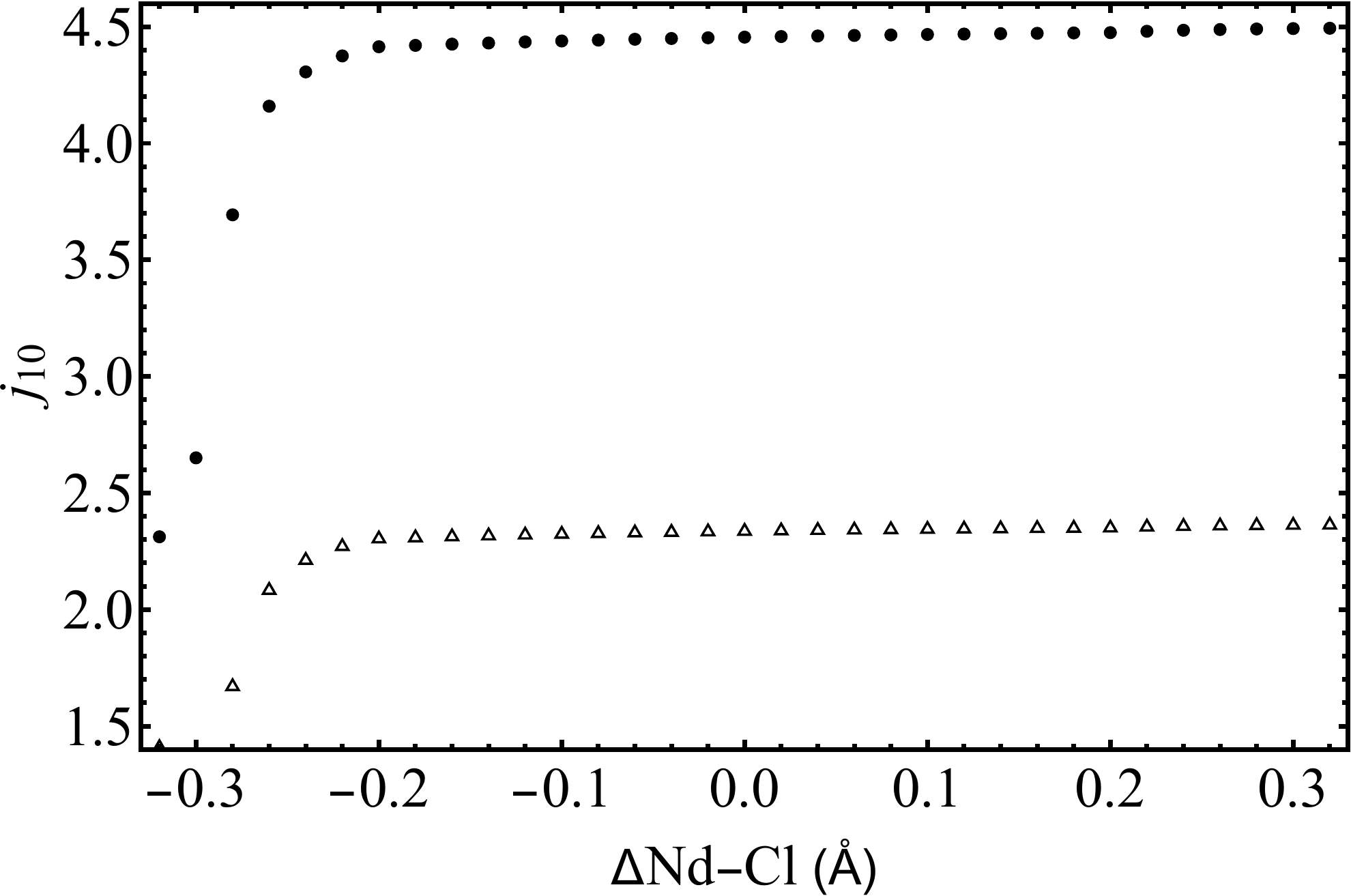}
\end{tabular}
\end{center}
\caption{
(a) $j_{10}(\alpha)$ for Nd (solid line) and Np (dashed line) clusters.
(b) $j_{10}$ of Nd cluster with respect to the totally symmetric deformation from the equilibrium Nd-Cl bond length, $\Delta$ Nd-Cl (\AA). 
The filled circles and open triangles indicate $j_{10}$ with $\alpha$ at the global maximum ($\alpha \approx \pi/2$) and the local maximum ($\alpha \approx -\pi/2$), respectively.
}
\label{Fig:j10}
\end{figure*}

\begin{table}[tb]
\caption{
$\alpha_0$ (rad), the total angular momentum $j_{kq}$ and crystal-field parameters $B_{k}$ (cm$^{-1}$) of Cs$_2$NaNdCl$_6$ and Cs$_2$ZrCl$_6$:Np$^{4+}$. (a) and (b) indicate the {\it ab initio} methodology (Sec. \ref{Sec:abinitio}).
}
\label{Table:B}
\begin{ruledtabular}
\begin{tabular}{cccccc}
& $k$ & $q$ & \multicolumn{2}{c}{Cs$_2$NaNdCl$_6$} & Cs$_2$ZrCl$_6$:Np$^{4+}$ \\
&     &     & (a) & (b) & (a) \\
\hline
$\alpha_0$   & &      & 1.620 & 1.614 & 1.631 \\
\hline
$j_{kq}$&1&0     & 4.455                  & 4.452 &  4.242 \\
        &3&0     & 6.88 $\times 10^{-3}$ & 9.28 $\times 10^{-3}$ &  2.68 $\times 10^{-2}$ \\
        &5&0     &$-3.15 \times 10^{-3}$ &$-4.34 \times 10^{-3}$ &$-1.10  \times 10^{-2}$ \\
        & &$\pm$4& 1.70 $\times 10^{-4}$ & 2.41 $\times 10^{-3}$ &$-4.54  \times 10^{-4}$ \\
        &7&0     & 6.03 $\times 10^{-4}$ & 6.80 $\times 10^{-4}$ &  1.25 $\times 10^{-3}$ \\
        & &$\pm$4&$-3.23 \times 10^{-4}$ & 2.15 $\times 10^{-4}$ &$-6.15  \times 10^{-4}$ \\
        &9&0     &$-3.76 \times 10^{-5}$ &$-4.71 \times 10^{-5}$ &$-1.12  \times 10^{-5}$ \\
        & &$\pm$4& 4.56 $\times 10^{-6}$ & 1.09 $\times 10^{-5}$ &  1.44 $\times 10^{-5}$ \\
        & &$\pm$8& 3.55 $\times 10^{-6}$ & 8.31 $\times 10^{-6}$ &  1.12 $\times 10^{-5}$ \\
%        &3&0     & 6.881 $\times 10^{-3}$ & 9.279 $\times 10^{-3}$ &  2.676 $\times 10^{-2}$ \\
%        &5&0     &$-3.154 \times 10^{-3}$ &$-4.340 \times 10^{-3}$ &$-1.098  \times 10^{-2}$ \\
%        & &$\pm$4& 1.698 $\times 10^{-4}$ & 2.405 $\times 10^{-3}$ &$-4.544  \times 10^{-4}$ \\
%        &7&0     & 6.031 $\times 10^{-4}$ & 6.800 $\times 10^{-4}$ &  1.253 $\times 10^{-3}$ \\
%        & &$\pm$4&$-3.226 \times 10^{-4}$ & 2.152 $\times 10^{-4}$ &$-6.149  \times 10^{-4}$ \\
%        &9&0     &$-3.764 \times 10^{-5}$ &$-4.714 \times 10^{-5}$ &$-1.122  \times 10^{-5}$ \\
%        & &$\pm$4& 4.556 $\times 10^{-6}$ & 1.085 $\times 10^{-5}$ &  1.441 $\times 10^{-5}$ \\
%        & &$\pm$8& 3.549 $\times 10^{-6}$ & 8.314 $\times 10^{-6}$ &  1.123 $\times 10^{-5}$ \\
\hline
$B_k$   &4&      & $-82.24$ & $-99.53$ & $-370.65$ \\
        &6&      &  $-8.65$ &  $-9.74$ &  $-39.73$ \\
        &8&      &    0.05  &    0.07  &   $-0.22$ 
\end{tabular}
\end{ruledtabular}
\end{table}

\subsubsection{{\it Ab initio} method}
\label{Sec:abinitio}
In order to obtain the electronic structure, embedded cluster calculations were performed with a post Hartree-Fock method. 
For the Cs$_2$NaNdCl$_6$ cluster, one Nd$^{3+}$ ion and the nearest eight Cl$^{-}$ ions are treated {\it ab initio}, and the distant atoms are replaced by point charges. 
The electronic structure was calculated using complete active space self-consistent field (CASSCF),
extended multi-state complete active space second-order perturbation theory 
(XMS-CASPT2) \cite{Granovsky2011, Shiozaki2011}, 
and spin-orbit restricted active space state interaction (SO-RASSI) methods with 
atomic-natural-orbital relativistic-correlation consistent-minimal basis (ANO-RCC-MB).
In the CASSCF calculations, 14 orbitals were included in the active space: $4f$ of the Nd$^{3+}$ ion
alongside with an additional set of seven $f$ functions (of the $5f$ kind of the metal site).
The dynamical electron correlation for these orbitals was taken into account within the XMS-CASPT2 approach.
The spin-orbit coupling was taken into account with SO-RASSI method, and the scalar relativistic effects were included in the basis set. 
The crystal-field states of Nd$^{3+}$ were calculated using two approaches: (a) CASSCF/SO-RASSI and (b) CASSCF/XMS-CASPT2/SO-RASSI.
All calculations were performed using Molcas 8 suite of programs \cite{Molcas8}.
The crystal-field states of Cs$_2$ZrCl$_6$:Np$^{4+}$ cluster within the same computational level were taken from the previous work \cite{Gamma8}.

\subsubsection{$\tilde{J}$-pseudospins of Cs$_2$NaNdCl$_6$ and Cs$_2$ZrCl$_6$:Np$^{4+}$}
\label{Sec:f3}
The calculated crystal-field levels of Cs$_2$NaNdCl$_6$ and Cs$_2$ZrCl$_6$:Np$^{4+}$ clusters are given in Table \ref{Table:E}.
In both cases, the irreducible representations of the crystal-field levels are $\Gamma_8$, $\Gamma_6$, $\Gamma_8$ in the order of increasing energy.
The obtained levels of Cs$_2$NaNdCl$_6$ are in good agreement with experimental data \cite{Zhou2006}, and the dynamical electron correlation makes the agreement better.
The {\it ab initio} $\Gamma$ multiplets were assigned by comparing the {\it ab initio} magnetic moment $\hat{\bm{\mu}}$ matrices and the structure of symmetry adapted model of $\hat{\bm{\mu}}$, which also enabled us to fix the relative phase factors. 

Following the method in Sec. \ref{Sec:Jpseudospin}, $\tilde{J}=9/2$ pseudospin states were defined. 
Figure \ref{Fig:j10}(a) shows the plot of $j_{10}(\alpha)$ 
as a function 
of $\alpha$, and the obtained $\alpha_0$, $j_{kq}(\alpha_0)$ and $B_k$ are listed in Table \ref{Table:B}.
The $\Gamma_8^{(1)}$ and $\Gamma_8^{(2)}$ states in Eq. (\ref{Eq:JM}) correspond to the excited and the ground $\Gamma_8$ multiplets, respectively. 
In order to check the principle 2, $j_{10}$ of the Nd cluster with respect to the strength of the crystal-field which is controlled by the totally symmetric displacements of ligand atoms. 
Fig. \ref{Fig:j10} (b) shows $j_{10}$ using two different $\alpha$:
one at the maximum point ($\alpha = \alpha_0 \approx \pi/2$) and the other at the second highest point ($\alpha \approx -\pi/2$) in Fig. \ref{Fig:j10}(a).
The first one (filled circle) continues to approach the atomic limit: $j_{10} = 4.494$ at the largest Nd-Cl. 
On the contrary, the second one (open triangle) remains of a much smaller value than the atomic one.
This demonstrates that the pseudospin states defined by the proposed algorithm indeed fulfills the two 
principles
outlined in Sec. \ref{Sec:pseudospin}.

The coefficients $j_{kq}$ in Table \ref{Table:B} shows that the first rank term in $\hat{J}_z$ is dominant, whereas the higher order terms are not negligible.
The discrepancy would be mainly explained by the covalency effect \cite{Ungur2017}.
The effect of covalency is seen by comparing Nd$^{3+}$ and Np$^{4+}$ ions:
due to the stronger delocalization of the $5f$ orbital in comparison with the $4f$ orbital, the bonding to the ligand becomes more important in the former, which results in a stronger reduction of $j_{10}$ in Np$^{4+}$ than in Nd$^{3+}$.
The discrepancy between the traditional crystal-field approach \cite{Lea1962} and the {\it ab initio} wave function based treatment described here also arises in the form of the crystal-field Hamiltonian, which involves as eighth-rank terms in the latter case.
We stress that the $\tilde{J}$-pseudospin Hamiltonian is more exact because, being derived directly from the {\it ab initio} electronic states, it reproduces by definition not only their energies but also all their electronic properties.

\section{Discussion}
\label{Sec:discussion}
The present $\tilde{J}$-pseudospin states fulfill both requirements presented in Sec. \ref{Sec:pseudospin}.
The same methodology will apply to other cases. 
For $\tilde{J} < 9/2$, the pseudospin states are uniquely defined by using the first principle as shown in Appendix \ref{A:JM}, whereas there are a few arbitrary parameters in the case of $\tilde{J} \ge 9/2$. 
The mixing parameters have to be introduced because some $\Gamma$ representations of the cubic group appear more than once under the descent of symmetry, $\tilde{J}\downarrow O_h$ (see Table \ref{Table:J_Gamma_Bk} and Appendix \ref{A:JM}).

One also should note that the present definition is one of the many equivalent definitions. 
In the case of octahedral systems, the eigenstates of $\hat{\mu}_z$ cannot be used as the pseudospin states which satisfy the symmetry requirements. 
This is explained by the fact that the applied magnetic field (Zeeman interaction) lowers the symmetry to and the eigenstates fulfill at most tetragonal symmetry.
Similar situation arises in all systems of cubic or icosahedral symmetry. 
In such cases, the idea of the approach proposed here should be applied. 
On the other hand, if the system has a low symmetry which 
in practice cannot be adiabatically changed into 
the cubic or higher one and the Zeeman interaction does not lower the symmetry, the conventional definition using the eigenstates of 
$\hat{\mu}_Z$ 
\cite{Ungur2017} will be reasonable. 

In Sec. \ref{Sec:Example}, to check the adiabatic connection between the obtained $\tilde{J}$-pseudospin states in cubic symmetry and atomic $J$-multiplets, {\it ab initio} calculations were performed at many cubic structures. 
However, this procedure could be significantly simplified by applying the indicator function approach proposed in Ref. \cite{Chibotaru2013}. 
With this method, the information of the atomic limit will be extracted from the wave function of the embedded system.

\section{Conclusions}
\label{Sec:conclusion}
In this work, the theory of $\tilde{J}$-pseudospin for cubic systems is developed. 
Using the symmetry, we derived the analytical expressions for all important $\tilde{J}$-pseudospin states.
Despite the high spatial and time-reversal symmetries, the large-$\tilde{J}$ pseudospin states cannot be completely determined due to the presence of the several arbitrary parameters. 
These free parameters are fixed by using the requirement of adiabatic connection.
In the case of $\tilde{J}$-pseudospin for the crystal-field model of $f$ elements, the free parameter is determined by maximizing the first rank term of total angular momentum because this definition allows $\tilde{J}$-pseudospin to converge to pure total angular momentum in the atomic limit. 
Although the original idea to fulfill the second requirement of the adiabatic connection is by performing many consecutive {\it ab initio} calculations varying the strength of interaction, the present algorithm enables us to determine the $\tilde{J}$-pseudospin based only on one calculation. 
With the derived $\tilde{J}$-pseudospin states, the total angular momentum and the crystal-field Hamiltonian contain terms of higher rank than fourth and sixth, which do not exist in the conventional model based on $f$-shells. 
The discrepancy can arise due to the effects which are not contained within the atomic shell model. 
Combining the developed approach and {\it ab initio} calculations, the crystal-field Hamiltonian of the Nd$^{3+}$ and Np$^{4+}$ ions in cubic environment were successfully derived. 
%The higher rank terms were also found to be non-negligible, which implies the importance of covalency effect in these systems.
Finally, we emphasize that the current methodology is not specific to the method for the calculations of wave functions, and is applicable to any multiplet states. 
Thus, with the increase of the accuracy of the {\it ab initio} calculations, %more precise pseudospins can be obtained. 
accurate definition of pseudospins can be achieved.

\section*{Acknowledgment}
N.I. was supported by Japan Society for the Promotion of Science Overseas Research Fellowship. % and L.U. was supported by FWO-Vlaanderen.

\appendix

\section{$\tilde{J}$ pseudospin states}
\label{A:JM}
The relation between the $\tilde{J}$ pseudospin states and $\Gamma$ crystal-field states,
\begin{eqnarray}
 \left(
  |\Gamma\gamma \rangle,
  |\Gamma'\gamma' \rangle,
  ...
 \right)
  = 
 \left(
 |M \rangle,
 |M' \rangle,
% |\tilde{J}M \rangle,
% |\tilde{J}M' \rangle,
 ...
 \right)
 U,
\end{eqnarray}
is derived up to $J=8$, where, $(|\Gamma\gamma \rangle, ...)$ and  $(|M \rangle, ...)$ are indices of crystal-field states and $\tilde{J}$-pseudospin states, respectively, $U$ is orthogonal matrix, and $|M\rangle$ stands for $|\tilde{J}M\rangle$.
The basis of the irreducible representations of cubic symmetry are taken from Ref. \cite{Koster1963}, and $|\tilde{J}M\rangle$ transform as spherical harmonics \cite{Varshalovich1988}.
The procedure of the derivation is similar to that of $\tilde{J}=9/2$ pseudospin states (Sec. \ref{Sec:pseudospin_cubic}).
The transformation coefficients $U$ between the non-repeating $\Gamma$ states (Table \ref{Table:J_Gamma_Bk}) and the $\tilde{J}$-pseudospin states are 
unambiguously
determined
by symmetry. 
The other $\Gamma$ states are determined up to their linear combinations, which are described by using the rotational matrices \cite{Varshalovich1988}:
\begin{eqnarray}
 R^{(2)}(\alpha) &=& 
 \begin{pmatrix}
  \cos \alpha & -\sin \alpha \\
  \sin \alpha &  \cos \alpha 
 \end{pmatrix},
\label{Eq:R2}
\end{eqnarray}
and
\begin{widetext}
\begin{eqnarray}
 R^{(3)}(\Omega) &=& 
 \begin{pmatrix}
  \cos \alpha \cos \beta \cos \gamma - \sin \alpha \sin \gamma & -\cos \alpha \cos \beta \sin \gamma - \sin \alpha \cos \gamma & \cos \alpha \sin \beta \\
  \sin \alpha \cos \beta \cos \gamma + \cos \alpha \sin \gamma & -\sin \alpha \cos \beta \sin \gamma + \cos \alpha \cos \gamma & \sin \alpha \sin \beta \\
  -\sin \beta \cos \gamma  & \sin \beta \sin \gamma & \cos \beta \\
 \end{pmatrix},
\label{Eq:R3}
\end{eqnarray}
where, $\alpha, \beta, \gamma$ are angles, and $\Omega = (\alpha, \beta, \gamma)$.
For the description of the $\tilde{J}$-pseudospin states of non-Kramers systems, symmetric and antisymmetric states are sometimes used: for positive $M = m$ ($m \le \tilde{J}$), 
\begin{eqnarray}
 |m_\pm \rangle &=& \frac{1}{\sqrt{2}} \left(|-m\rangle \pm |+m\rangle\right).
\label{Eq:mpm}
\end{eqnarray}

\subsection{Non-Kramers ion}

\subsubsection{$\tilde{J} = 2$}
%The relation between the crystal-field and the pseudospin states are determined by symmetry:
\begin{eqnarray}
|\Gamma_3\theta\rangle &=& |0\rangle, 
\quad
|\Gamma_3\epsilon\rangle = |2_+\rangle,  
\quad
%\nonumber\\
|\Gamma_5,0\rangle = |2_-\rangle,  
\quad
|\Gamma_5,\mp 1\rangle = \pm |\pm 1\rangle.
%%% TYPE 1
%|\Gamma_3\theta\rangle &=& |0\rangle, 
%\quad
%|\Gamma_3\epsilon\rangle = |2_+\rangle,  
%%\quad
%\nonumber\\
%|\Gamma_5,0\rangle &=& |2_-\rangle,  
%\quad
%|\Gamma_5,\mp 1\rangle = \pm |\pm 1\rangle.
%%% TYPE 2
%\left(
%|\Gamma_3\theta\rangle, |\Gamma_3\epsilon\rangle,
%|\Gamma_5,0\rangle, |\Gamma_5,-1\rangle, |\Gamma_5,+1\rangle
%\right)
%\nonumber\\
%\left(
% |-2\rangle, |-1\rangle, |0\rangle, |+1\rangle, |+2\rangle
%\right) U, 
%\end{eqnarray}
%where, 
%\begin{eqnarray}
%U &=& 
%\begin{pmatrix}
% 0 & \frac{1}{\sqrt{2}} & \frac{1}{\sqrt{2}} & 0 & 0 \\
% 0 & 0 & 0 & 0 & -1\\
% 1 & 0 & 0 & 0 & 0 \\
% 0 & 0 & 0 & 1 & 0 \\
% 0 & \frac{1}{\sqrt{2}} & -\frac{1}{\sqrt{2}} & 0 & 0 \\
%\end{pmatrix}
\end{eqnarray}
The crystal-field parameter $B_4$ is given by 
\begin{eqnarray}
 B_4 &=& \frac{E_{\Gamma_3} - E_{\Gamma_5}}{10}.
\end{eqnarray}

\subsubsection{$\tilde{J} = 3$}
%The relation between the crystal-field and the pseudospin states are determined by symmetry:
\begin{eqnarray}
\left(
|\Gamma_2\rangle,
|\Gamma_4,0\rangle,
|\Gamma_4,-1\rangle,
|\Gamma_4,+1\rangle,
|\Gamma_5,0\rangle,
|\Gamma_5,-1\rangle,
|\Gamma_5,+1\rangle
\right)
\nonumber\\
= 
\left(
|2_-\rangle,  
|0\rangle , 
|2_+\rangle,  
|-1\rangle, 
|+3\rangle, 
|+1\rangle, 
|-3\rangle
\right)
%\nonumber\\
%\end{eqnarray}
%where,
%\begin{eqnarray}
% U &=& 
\nonumber\\
 \times
 \begin{pmatrix}
  1 & 0 & 0 & 0 & 0 & 0 & 0 \\
  0 & 1 & 0 & 0 & 0 & 0 & 0 \\
  0 & 0 & 0 & 0 & 1 & 0 & 0 \\
  0 & 0 & M_1 & 0 & 0 & 0 & M_2 \\
  0 & 0 & 0 & M_1 & 0 & M_2 & 0 \\
%  0 & 0 & -\frac{1}{2}\sqrt{\frac{3}{2}} & 0 & 0 & 0 &  \frac{1}{2}\sqrt{\frac{5}{2}} \\
%  0 & 0 & -\frac{1}{2}\sqrt{\frac{5}{2}} & 0 & 0 & 0 & -\frac{1}{2}\sqrt{\frac{3}{2}} \\
%  0 & 0 & 0 & -\frac{1}{2}\sqrt{\frac{3}{2}} & 0 &  \frac{1}{2}\sqrt{\frac{5}{2}} & 0 \\
%  0 & 0 & 0 & -\frac{1}{2}\sqrt{\frac{5}{2}} & 0 & -\frac{1}{2}\sqrt{\frac{3}{2}} & 0 \\
 \end{pmatrix},
\label{Eq:UJ3}
\end{eqnarray}
where, 
\begin{eqnarray}
 M_1 &=& 
 \begin{pmatrix}
   -\frac{1}{2}\sqrt{\frac{3}{2}}  \\
   -\frac{1}{2}\sqrt{\frac{5}{2}}  \\
 \end{pmatrix},
\quad
 M_2 = 
 \begin{pmatrix}
   \frac{1}{2}\sqrt{\frac{5}{2}} \\
  -\frac{1}{2}\sqrt{\frac{3}{2}} \\
 \end{pmatrix}.
\end{eqnarray}
The crystal-field parameters are 
\begin{eqnarray}
 B_4 &=& \frac{-6E_{\Gamma_2} + 9E_{\Gamma_4} - 3E_{\Gamma_5}}{44}, 
%\nonumber\\
\quad
 B_6 = \frac{-4E_{\Gamma_2} -5E_{\Gamma_4} + 9E_{\Gamma_5}}{616}.
\end{eqnarray}

\subsubsection{$\tilde{J} = 4$}
%The relation between the crystal-field and the pseudospin states are determined by symmetry:
\begin{eqnarray}
\left(
|\Gamma_1\rangle,
|\Gamma_3\theta\rangle,
|\Gamma_3\epsilon\rangle,
|\Gamma_4,0\rangle,
|\Gamma_4,-1\rangle,
|\Gamma_4,+1\rangle,
|\Gamma_5,0\rangle,
|\Gamma_5,-1\rangle,
|\Gamma_5,+1\rangle
\right) 
\nonumber\\
= 
\left(
|4_+\rangle, 
|0\rangle , 
|2_+\rangle,  
|4_-\rangle, 
|2_-\rangle,  
|-1\rangle, 
|+3\rangle, 
|+1\rangle, 
|-3\rangle
\right)
%\end{eqnarray}
%where,
%\begin{eqnarray}
% U &=& 
\nonumber\\
 \times
 \begin{pmatrix}
  \frac{1}{2}\sqrt{\frac{5}{3}} & -\frac{1}{2}\sqrt{\frac{7}{3}} & 0 & 0 & 0 & 0 & 0 & 0 & 0 \\
  \frac{1}{2}\sqrt{\frac{7}{3}} &  \frac{1}{2}\sqrt{\frac{5}{3}} & 0 & 0 & 0 & 0 & 0 & 0 & 0 \\
  0 & 0 & -1 & 0 & 0 & 0 & 0 & 0 & 0 \\
  0 & 0 & 0 & 1 & 0 & 0 & 0 & 0 & 0 \\
  0 & 0 & 0 & 0 & 0 & 0 & 1 & 0 & 0 \\
  0 & 0 & 0 & 0 & -M_1 & 0 & 0 & 0 & -M_2 \\
  0 & 0 & 0 & 0 & 0 & M_1 & 0 & M_2 & 0  \\
%  0 & 0 & 0 & 0 & -\frac{1}{2}\sqrt{\frac{7}{2}} & 0 & 0 & 0 & \frac{1}{2\sqrt{2}}\\
%  0 & 0 & 0 & 0 & -\frac{1}{2\sqrt{2}} & 0 & 0 & 0 & -\frac{1}{2}\sqrt{\frac{7}{2}}\\
%  0 & 0 & 0 & 0 & 0 & \frac{1}{2}\sqrt{\frac{7}{2}} & 0 & -\frac{1}{2\sqrt{2}} & 0  \\
%  0 & 0 & 0 & 0 & 0 & \frac{1}{2\sqrt{2}} & 0 & \frac{1}{2}\sqrt{\frac{7}{2}} & 0 \\
 \end{pmatrix},
\label{Eq:UJ4}
\end{eqnarray}
where, 
\begin{eqnarray}
 M_1 &=& 
 \frac{1}{2\sqrt{2}}
 \begin{pmatrix}
  \sqrt{7} \\
  1 
 \end{pmatrix},
\quad
 M_2 =
 \frac{1}{2\sqrt{2}}
 \begin{pmatrix}
  -1 \\
  \sqrt{7} 
 \end{pmatrix}.
\end{eqnarray}
The crystal-field parameters are given by 
\begin{eqnarray}
 B_4 &=& \frac{7(14E_{\Gamma_1} + 4E_{\Gamma_3} + 21E_{\Gamma_4} - 39E_{\Gamma_5})}{858},
\quad
 B_6 = \frac{-20E_{\Gamma_1} + 32 E_{\Gamma_3} + 3(E_{\Gamma_4} - 5E_{\Gamma_5})}{990},
\nonumber\\
 B_8 &=& \frac{5E_{\Gamma_1} + 7E_{\Gamma_3} - 12E_{\Gamma_4}}{1560}.
% B_4 &=& \frac{7(14E_{\Gamma_1} + 4E_{\Gamma_3} + 21E_{\Gamma_4} - 39E_{\Gamma_5})}{858},
%\nonumber\\
% B_6 &=& \frac{-20E_{\Gamma_1} + 32 E_{\Gamma_3} + 3(E_{\Gamma_4} - 5E_{\Gamma_5})}{990},
%\nonumber\\
% B_8 &=& \frac{5E_{\Gamma_1} + 7E_{\Gamma_3} - 12E_{\Gamma_4}}{1560}.
\end{eqnarray}

\subsubsection{$\tilde{J} = 5$}
%The crystal states are constructed from $\tilde{J}$ pseudospin states,
\begin{eqnarray}
\left(
|\Gamma_3\theta\rangle,
|\Gamma_3\epsilon\rangle,
|\Gamma_4^{(1)},0\rangle,
|\Gamma_4^{(2)},0\rangle,
|\Gamma_4^{(1)},-1\rangle,
|\Gamma_4^{(2)},-1\rangle,
|\Gamma_4^{(1)},+1\rangle,
|\Gamma_4^{(2)},+1\rangle,
|\Gamma_5,0\rangle,
|\Gamma_5,-1\rangle,
|\Gamma_5,+1\rangle
\right)
\nonumber\\
=
\left(
|4_-\rangle, 
|2_-\rangle,  
|4_+\rangle, 
|0\rangle , 
|2_+\rangle,  
|-5\rangle, 
|-1\rangle, 
|+3\rangle, 
|+5\rangle, 
|+1\rangle, 
|-3\rangle
\right)
%\end{eqnarray}
%by using 
%\begin{eqnarray}
% U &=& 
\nonumber\\
\times
 \begin{pmatrix}
   1 & 0 & 0 & 0 & 0 & 0 & 0 & 0 \\
   0 & -1 & 0 & 0 & 0 & 0 & 0 & 0 \\
   0 & 0 & R^{(2)}(\alpha) & 0 & 0 & 0 & 0 & 0 \\
   0 & 0 & 0 & 0 & 0 & 1 & 0 & 0 \\
   0 & 0 & 0 & M_1 R^{(2)}(\alpha) & 0 & 0 & 0 & M_2 \\
   0 & 0 & 0 & 0 & M_1 R^{(2)}(\alpha) & 0 & M_2 & 0 \\
 \end{pmatrix},
\label{Eq:UJ5}
\end{eqnarray}
where, $M_1$ and $M_2$ are defined by 
\begin{eqnarray}
 M_1 = 
 \frac{1}{8\sqrt{2}} 
 \begin{pmatrix}
   \sqrt{5}  & 3\sqrt{7} \\
   \sqrt{42} & \sqrt{30} \\
   -9        & \sqrt{35} 
 \end{pmatrix},
\quad
 M_2 = 
 \frac{1}{4\sqrt{2}} 
 \begin{pmatrix}
  \sqrt{15} \\
  -\sqrt{14} \\
 -\sqrt{3} 
 \end{pmatrix}.
\end{eqnarray}
%Since both $\Gamma_3$ and $\Gamma_5$ states appear once, the matrix elements for these states are determined by symmetry. 
%On the other hand, $\Gamma_4$ states appear twice. 
%In order to describe two sets of $\Gamma_4$ states from two $\tilde{J}$ pseudospin states, arbitrary angle $\alpha$ is introduced. 

\subsubsection{$\tilde{J} = 6$}
%The crystal states are constructed from $\tilde{J}$ pseudospin states,
\begin{eqnarray}
\left(
|\Gamma_1\rangle,
|\Gamma_2\rangle,
|\Gamma_3\theta\rangle,
|\Gamma_3\epsilon\rangle,
|\Gamma_4,0\rangle,
|\Gamma_4,-1\rangle,
|\Gamma_4,+1\rangle,
|\Gamma_5^{(1)},0\rangle,
|\Gamma_5^{(2)},0\rangle,
|\Gamma_5^{(1)},-1\rangle,
|\Gamma_5^{(2)},-1\rangle,
|\Gamma_5^{(1)},+1\rangle,
|\Gamma_5^{(2)},+1\rangle
\right)
\nonumber\\
 = 
\left(
|4_+\rangle, 
|0\rangle , 
|6_+\rangle, 
|2_+\rangle,  
|4_-\rangle, 
|6_-\rangle, 
|2_-\rangle,  
|-5\rangle, 
|-1\rangle, 
|+3\rangle, 
|+5\rangle, 
|+1\rangle, 
|-3\rangle 
\right) 
\nonumber\\
\times
%\end{eqnarray}
%using 
%\begin{eqnarray}
% U &=& 
 \begin{pmatrix}
  \frac{1}{2} \sqrt{\frac{7}{2}} & 0 & \frac{1}{2\sqrt{2}} & 0 & 0 & 0 & 0 & 0 & 0 & 0 \\
 -\frac{1}{2\sqrt{2}} & 0 & \frac{1}{2} \sqrt{\frac{7}{2}} & 0 & 0 & 0 & 0 & 0 & 0 & 0 \\
  0 &  \frac{\sqrt{5}}{4}   & 0 & \frac{\sqrt{11}}{4} & 0 & 0& 0 & 0 & 0 & 0 \\
  0 & -\frac{\sqrt{11}}{4}  & 0  & \frac{\sqrt{5}}{4} & 0 & 0& 0 & 0 & 0 & 0 \\
  0 & 0 & 0 & 0 & 1 & 0 & 0 & 0 & 0 & 0 \\
  0 & 0 & 0 & 0 & 0 & 0 & 0 & R^{(2)}(\alpha) & 0 & 0 \\
  0 & 0 & 0 & 0 & 0 & M_1 & 0 & 0 & 0 & M_2 R^{(2)}(\alpha) \\
  0 & 0 & 0 & 0 & 0 & 0 & -M_1 & 0 & -M_2 R^{(2)}(\alpha) & 0 \\
 \end{pmatrix},
\label{Eq:UJ6}
\end{eqnarray}
where, %$M_1$ and $M_2$ are given as
\begin{eqnarray}
 M_1 = 
 \frac{1}{4\sqrt{2}}
 \begin{pmatrix}
   -\sqrt{11} \\
   \sqrt{6} \\
   -\sqrt{15} \\
 \end{pmatrix},
\quad
 M_2 = 
 \frac{1}{16}
 \begin{pmatrix}
  \sqrt{3} & \sqrt{165} \\
  3\sqrt{22} & \sqrt{10} \\
  \sqrt{55} & -9 
 \end{pmatrix}.
\end{eqnarray}

\subsubsection{$\tilde{J} = 7$}
%The transformation from $\tilde{J}=7$ pseudospin states to the crystal-field states are 
\begin{eqnarray}
\left(
|\Gamma_2\rangle,
|\Gamma_3\theta\rangle,
|\Gamma_3\epsilon\rangle,
|\Gamma_4^{(1)},0\rangle,
|\Gamma_4^{(2)},0\rangle,
|\Gamma_4^{(1)},-1\rangle,
|\Gamma_4^{(2)},-1\rangle,
|\Gamma_4^{(1)},+1\rangle,
|\Gamma_4^{(2)},+1\rangle,
|\Gamma_5^{(1)},0\rangle,
|\Gamma_5^{(2)},0\rangle,
\right.
\nonumber\\
\left.
|\Gamma_5^{(1)},-1\rangle,
|\Gamma_5^{(2)},-1\rangle,
|\Gamma_5^{(1)},+1\rangle,
|\Gamma_5^{(2)},+1\rangle
\right)
\nonumber\\
= 
\left(
|6_-\rangle, 
|2_-\rangle,  
|4_-\rangle, 
|4_+\rangle, 
|0\rangle , 
|6_+\rangle, 
|2_+\rangle,  
|-5\rangle, 
|-1\rangle, 
|+3\rangle, 
|+7\rangle,  
|+5\rangle, 
|+1\rangle, 
|-3\rangle, 
|-7\rangle
\right) 
%\end{eqnarray}
%where, $U$ is defined by 
%\begin{eqnarray}
% U &=& 
\nonumber\\
 \times
 \begin{pmatrix}
   \frac{1}{2}\sqrt{\frac{11}{6}} & 0 & -\frac{1}{2}\sqrt{\frac{13}{6}}  & 0 & 0 & 0 & 0 & 0 & 0 \\
   \frac{1}{2}\sqrt{\frac{13}{6}} & 0 &  \frac{1}{2}\sqrt{\frac{11}{6}}  & 0 & 0 & 0 & 0 & 0 & 0 \\
   0 & 1 & 0 & 0 & 0 & 0 & 0 & 0 & 0 \\
   0 & 0 & 0 & R^{(2)}(\alpha) & 0 & 0 & 0 & 0 & 0 \\
   0 & 0 & 0 & 0 & 0 & 0 & R^{(2)}(\beta) & 0 & 0 \\
   0 & 0 & 0 & 0 & M_1 R^{(2)}(\alpha) & 0 & 0 & 0 & M_2 R^{(2)}(\beta) \\ 
   0 & 0 & 0 & 0 & 0 & M_1 R^{(2)}(\alpha) & 0 & M_2 R^{(2)}(\beta) & 0 
 \end{pmatrix},
\label{Eq:UJ7}
\end{eqnarray}
where, 
\begin{eqnarray}
 M_1 = 
 \frac{1}{32}
 \begin{pmatrix}
  25 & -\sqrt{231} \\
 -3\sqrt{33} & -5\sqrt{7} \\
  \sqrt{11} & -3\sqrt{21} \\
  -\sqrt{91} & -\sqrt{429} 
 \end{pmatrix},
\quad
 M_2 = 
 \frac{1}{32\sqrt{2}}
 \begin{pmatrix}
   5\sqrt{13}  & \sqrt{11} \\
   \sqrt{429}  & 15\sqrt{3} \\
  -3\sqrt{143} & 19 \\
  -\sqrt{7}    & -\sqrt{1001} \\
 \end{pmatrix}.
\end{eqnarray}

\subsubsection{$\tilde{J} = 8$}
%The transformation from $\tilde{J}=8$ pseudospin states to the crystal-field states is given as 
\begin{eqnarray}
\left(
|\Gamma_1\rangle,
|\Gamma_3^{(1)}\theta\rangle,
|\Gamma_3^{(2)}\theta\rangle,
|\Gamma_3^{(1)}\epsilon\rangle,
|\Gamma_3^{(2)}\epsilon\rangle, 
|\Gamma_4^{(1)},0\rangle,
|\Gamma_4^{(2)},0\rangle,
|\Gamma_4^{(1)},-1\rangle,
|\Gamma_4^{(2)},-1\rangle,
|\Gamma_4^{(1)},+1\rangle,
|\Gamma_4^{(2)},+1\rangle,
|\Gamma_5^{(1)},0\rangle,
|\Gamma_5^{(2)},0\rangle,
\right.
\nonumber\\
\left.
|\Gamma_5^{(1)},-1\rangle,
|\Gamma_5^{(2)},-1\rangle,
|\Gamma_5^{(1)},+1\rangle,
|\Gamma_5^{(2)},+1\rangle
\right)
\nonumber\\
=
\left(
|8_+\rangle,
|4_+\rangle,
|0\rangle ,
|6_+\rangle,
|2_+\rangle, 
|8_-\rangle,
|4_-\rangle,
|6_-\rangle,
|2_-\rangle, 
|-5\rangle,
|-1\rangle,
|+3\rangle,
|+7\rangle, 
|+5\rangle,
|+1\rangle,
|-3\rangle,
|-7\rangle
\right) 
\nonumber\\
%\end{eqnarray}
%where, 
%\begin{eqnarray}
\times
 \begin{pmatrix}
  M_1 & M_2 R^{(2)}(\alpha) & 0 & 0 & 0 & 0 & 0 & 0 & 0 \\
  0 &  0 & M_3 M_2 R^{(2)}(\alpha) & 0 & 0 & 0 & 0 & 0 & 0 \\
  0 & 0 & 0 & R^{(2)}(\beta) & 0 & 0 & 0 & 0 & 0 \\
  0 & 0 & 0 & 0 & 0 & 0 & R^{(2)}(\gamma) & 0 & 0 \\
  0 & 0 & 0 & 0 & -M_4 R^{(2)}(\beta) & 0 & 0 & 0 & -M_5 R^{(2)}(\gamma) \\ 
  0 & 0 & 0 & 0 & 0 & M_4 R^{(2)}(\beta) & 0 & M_5 R^{(2)}(\gamma) & 0 \\
 \end{pmatrix},
\nonumber\\
\label{Eq:UJ8}
\end{eqnarray}
where, 
%Here, matrices $M_n$ ($n=1,2,3,4,5$) are, respectively, 
\begin{eqnarray}
 M_1 &=&
 \frac{1}{8\sqrt{3}}
 \begin{pmatrix}
  \sqrt{65} \\
  2\sqrt{7} \\
  3\sqrt{11} 
 \end{pmatrix},
\quad
 M_2 = 
 \frac{1}{8\sqrt{93}}
 \begin{pmatrix}
  \sqrt{2145} & -16 \sqrt{7} \\
  2\sqrt{231} & 8\sqrt{65} \\
  -31\sqrt{3} & 0 
 \end{pmatrix},
\quad
 M_3 = 
 \frac{1}{96}
 \begin{pmatrix}
   3\sqrt{10} &  6\sqrt{182} & -3\sqrt{286} \\
  \sqrt{6006} & -2\sqrt{330} & -3\sqrt{210} \\
%   \sqrt{30} &  2\sqrt{546} & -\sqrt{858} \\
% \sqrt{2002} & -2\sqrt{101} & -3\sqrt{70}
 \end{pmatrix},
\nonumber\\
 M_4 &=&
 \frac{1}{32}
 \begin{pmatrix}
   \sqrt{35}  & 3\sqrt{13} \\
   \sqrt{715} & \sqrt{77} \\
   \sqrt{273} & -5\sqrt{15} \\
   1 & \sqrt{455} 
 \end{pmatrix},
\quad
 M_5 = 
 \frac{1}{32\sqrt{2}}
 \begin{pmatrix}
  7\sqrt{21}  & -\sqrt{715} \\
 -\sqrt{429}  & -\sqrt{35} \\
  \sqrt{455}  & 3\sqrt{33} \\
  3\sqrt{15}  & \sqrt{1001} 
 \end{pmatrix}.
\nonumber\\
\end{eqnarray}
%Thus, the pseudospin states contain three angles $\alpha, \beta, \gamma$.

\subsection{Kramers ion}
\subsubsection{$\tilde{J} = 5/2$}
%The transformation matrix from the $\tilde{J}=5/2$ pseudospin states 
\begin{eqnarray}
\left(
 \left|\Gamma_7, -\frac{1}{2} \right\rangle,
 \left|\Gamma_7, +\frac{1}{2} \right\rangle,
 \left|\Gamma_8, -\frac{3}{2} \right\rangle,
 \left|\Gamma_8, +\frac{3}{2} \right\rangle,
 \left|\Gamma_8, -\frac{1}{2} \right\rangle,
 \left|\Gamma_8, +\frac{1}{2} \right\rangle
\right)
\nonumber\\
 = 
\left(
 \left|-\frac{5}{2}\right\rangle, 
 \left|+\frac{3}{2}\right\rangle, 
 \left|+\frac{5}{2}\right\rangle, 
 \left|-\frac{3}{2}\right\rangle,
 \left|-\frac{1}{2}\right\rangle, 
 \left|+\frac{1}{2}\right\rangle
\right) 
%\end{eqnarray}
%where, $U$ is defined by 
%\begin{eqnarray}
% U &=& 
%\nonumber\\
% \times
 \begin{pmatrix}
  M_1 & 0 & 0 & M_2 & 0 & 0 \\   
  0 & M_1 & M_2 & 0 & 0 & 0 \\
%   \frac{1}{\sqrt{6}} & 0 & 0 & \sqrt{\frac{5}{6}} & 0 & 0 \\
%  -\sqrt{\frac{5}{6}} & 0 & 0 & \frac{1}{\sqrt{6}} & 0 & 0 \\   
%  0 &  \frac{1}{\sqrt{6}} & \sqrt{\frac{5}{6}} & 0 & 0 & 0 \\
%  0 & -\sqrt{\frac{5}{6}} & \frac{1}{\sqrt{6}} & 0 & 0 & 0 \\
  0 & 0 & 0 & 0 & 1 & 0 \\  
  0 & 0 & 0 & 0 & 0 & 1 \\  
 \end{pmatrix},
\end{eqnarray}
where, 
\begin{eqnarray}
 M_1 &=& 
 \begin{pmatrix}
   \frac{1}{\sqrt{6}} \\
  -\sqrt{\frac{5}{6}} \\   
 \end{pmatrix},
\quad
 M_2 = 
 \begin{pmatrix}
 \sqrt{\frac{5}{6}} \\
 \frac{1}{\sqrt{6}} \\
 \end{pmatrix}.
\end{eqnarray}
The crystal-field parameter is given by 
\begin{eqnarray}
 B_4 &=& -\frac{E_{\Gamma_7} - E_{\Gamma_8}}{6}.
\end{eqnarray}

%The $\tilde{J} = 5/2$ pseudospin states are determined by symmetry:
%\begin{eqnarray}
% \left|\tilde{J}, \mp \frac{5}{2}\right\rangle &=& \frac{1}{\sqrt{6}} \left|\Gamma_7, \mp \frac{1}{2} \right\rangle 
%                                               \mp \sqrt{\frac{5}{6}} \left|\Gamma_8, \pm \frac{3}{2} \right\rangle, 
%\nonumber\\
% \left|\tilde{J}, \mp \frac{3}{2}\right\rangle &=& -\sqrt{\frac{5}{6}} \left|\Gamma_7, \pm \frac{1}{2} \right\rangle
%                                               \pm \frac{1}{\sqrt{6}}  \left|\Gamma_8, \mp \frac{3}{2} \right\rangle,
%\nonumber\\
% \left|\tilde{J}, \mp \frac{1}{2}\right\rangle &=& \mp \left|\Gamma_8, \mp \frac{1}{2} \right\rangle.
%\label{Eq:J52}
%\end{eqnarray}
%The crystal-field contains up to fourth rank term and the parameters are given by 
%\begin{eqnarray}
% %B_0 = 2E_{\Gamma_7} + 4E_{\Gamma_8}, \quad 
% B_4 = -\frac{14}{3}(E_{\Gamma_7} - E_{\Gamma_8}).
%\label{Eq:CF52}
%\end{eqnarray}

%\begin{widetext}
\subsubsection{$\tilde{J} = 7/2$}
%The transformation matrix from the $\tilde{J}=7/2$ pseudospin states 
\begin{eqnarray}
\left(
 \left|\Gamma_6, -\frac{1}{2} \right\rangle,
 \left|\Gamma_6, +\frac{1}{2} \right\rangle,
 \left|\Gamma_7, -\frac{1}{2} \right\rangle,
 \left|\Gamma_7, +\frac{1}{2} \right\rangle,
 \left|\Gamma_8, -\frac{1}{2} \right\rangle,
 \left|\Gamma_8, +\frac{1}{2} \right\rangle,
 \left|\Gamma_8, -\frac{3}{2} \right\rangle,
 \left|\Gamma_8, +\frac{3}{2} \right\rangle
\right)
\nonumber\\
 = 
\left(
 \left|+\frac{7}{2}\right\rangle, 
 \left|-\frac{1}{2}\right\rangle, 
 \left|-\frac{7}{2}\right\rangle, 
 \left|+\frac{1}{2}\right\rangle, 
 \left|-\frac{5}{2}\right\rangle, 
 \left|+\frac{3}{2}\right\rangle, 
 \left|+\frac{5}{2}\right\rangle, 
 \left|-\frac{3}{2}\right\rangle
\right)
%\end{eqnarray}
%where, $U$ is 
%\begin{eqnarray}
% U &=& 
\nonumber\\
\times
 \begin{pmatrix} 
%  \frac{1}{2} \sqrt{\frac{5}{3}} & 0 & 0 & 0 & -\frac{1}{2} \sqrt{\frac{7}{3}} & 0 & 0 & 0 \\
%  \frac{1}{2} \sqrt{\frac{7}{3}} & 0 & 0 & 0 &  \frac{1}{2} \sqrt{\frac{5}{3}} & 0 & 0 & 0 \\
%  0 & -\frac{1}{2} \sqrt{\frac{5}{3}} & 0 & 0 & 0 & -\frac{1}{2} \sqrt{\frac{7}{3}} & 0 & 0 \\
%  0 & -\frac{1}{2} \sqrt{\frac{7}{3}} & 0 & 0 & 0 &  \frac{1}{2} \sqrt{\frac{5}{3}} & 0 & 0 \\
%  0 & 0 & -\frac{\sqrt{3}}{2} & 0 & 0 & 0 & 0 & -\frac{1}{2}        \\
%  0 & 0 &  \frac{1}{2} & 0       & 0 & 0 & 0 & -\frac{\sqrt{3}}{2} \\
%  0 & 0 & 0 &  \frac{\sqrt{3}}{2}& 0 & 0 & -\frac{1}{2} & 0  \\
%  0 & 0 & 0 & -\frac{1}{2}        & 0 & 0 & -\frac{\sqrt{3}}{2} & 0 \\
  M_1 & 0 & 0 & 0 & M_3 & 0 & 0 & 0 \\
  0 & -M_1 & 0 & 0 & 0 & M_3 & 0 & 0 \\
  0 & 0 &  M_2 & 0 & 0 & 0 & 0 & M_4 \\
  0 & 0 & 0 & -M_2 & 0 & 0 & M_4 & 0  \\
 \end{pmatrix}, 
\end{eqnarray}
where, 
\begin{eqnarray}
 M_1 &=& 
 \begin{pmatrix}
  \frac{1}{2} \sqrt{\frac{5}{3}} \\
  \frac{1}{2} \sqrt{\frac{7}{3}} \\
 \end{pmatrix},
\quad
 M_2 = 
 \begin{pmatrix}
 -\frac{\sqrt{3}}{2} \\
  \frac{1}{2}        \\
 \end{pmatrix},
\quad
 M_3 = 
 \begin{pmatrix}
   -\frac{1}{2} \sqrt{\frac{7}{3}}  \\
    \frac{1}{2} \sqrt{\frac{5}{3}}  \\
 \end{pmatrix},
\quad
 M_4 = 
 \begin{pmatrix}
 -\frac{1}{2}        \\
 -\frac{\sqrt{3}}{2} \\
 \end{pmatrix}.
\end{eqnarray}
%The $\tilde{J} = 7/2$ pseudospin states are also determined by symmetry:
%\begin{eqnarray}
% \left|\tilde{J}, \mp \frac{7}{2}\right\rangle &=& \pm \frac{1}{2} \sqrt{\frac{5}{3}} \left|\Gamma_6, \pm \frac{1}{2} \right\rangle
%                                                     + \frac{1}{2} \sqrt{\frac{7}{3}}   \left|\Gamma_8, \pm \frac{1}{2} \right\rangle,
%\nonumber\\
% \left|\tilde{J}, \mp \frac{5}{2}\right\rangle &=& \mp \ \frac{\sqrt{3}}{2} \left|\Gamma_7, \mp \frac{1}{2} \right\rangle
%                                                     + \frac{1}{2} \left|\Gamma_8, \pm \frac{3}{2} \right\rangle,
%\nonumber\\
% \left|\tilde{J}, \mp \frac{3}{2}\right\rangle &=& \mp \frac{1}{2} \left|\Gamma_7, \pm \frac{1}{2} \right\rangle
%                                                     + \frac{\sqrt{3}}{2} \left|\Gamma_8, \mp \frac{3}{2} \right\rangle,
%\nonumber\\
% \left|\tilde{J}, \mp \frac{1}{2}\right\rangle &=& \mp \frac{1}{2} \sqrt{\frac{7}{3}} \left|\Gamma_6, \mp \frac{1}{2} \right\rangle
%                                                     - \frac{1}{2} \sqrt{\frac{5}{3}} \left|\Gamma_8, \mp \frac{1}{2} \right\rangle. 
%\label{Eq:J72}
%\end{eqnarray}
%The crystal-field contains up to sixth rank terms and the parameters are given by 
%\begin{eqnarray}
%% B_0 &=& 2E_{\Gamma_6} + 2E_{\Gamma_7} + 4E_{\Gamma_8}, 
%%\nonumber\\
% B_4& =& \frac{1}{3}(7E_{\Gamma_6} - 9E_{\Gamma_7} + 2E_{\Gamma_8}),
%\nonumber\\
% B_6 &=& -5E_{\Gamma_6} -3E_{\Gamma_7} + 8E_{\Gamma_8}.
%\label{Eq:CF72}
%\end{eqnarray}
The crystal-field parameters are calculated as 
\begin{eqnarray}
 B_4 = \frac{49E_{\Gamma_6} - 63 E_{\Gamma_7} + 14 E_{\Gamma_8}}{264},
\quad
 B_6 = \frac{-5E_{\Gamma_6} - 3E_{\Gamma_7} + 8E_{\Gamma_8}}{264}.
\end{eqnarray}

\subsubsection{$\tilde{J} = 9/2$}
%The transformation matrix from the $\tilde{J}=9/2$ pseudospin states 
\begin{eqnarray}
\left(
 \left|\Gamma_6, -\frac{1}{2} \right\rangle,
 \left|\Gamma_6, +\frac{1}{2} \right\rangle,
 \left|\Gamma_8^{(1)}, -\frac{1}{2} \right\rangle,
 \left|\Gamma_8^{(2)}, -\frac{1}{2} \right\rangle,
 \left|\Gamma_8^{(1)}, +\frac{1}{2} \right\rangle,
 \left|\Gamma_8^{(2)}, +\frac{1}{2} \right\rangle,
 \left|\Gamma_8^{(1)}, -\frac{3}{2} \right\rangle,
 \left|\Gamma_8^{(2)}, -\frac{3}{2} \right\rangle,
\right.
\nonumber\\
\left.
 \left|\Gamma_8^{(1)}, +\frac{3}{2} \right\rangle,
 \left|\Gamma_8^{(2)}, +\frac{3}{2} \right\rangle
\right)
\nonumber\\
 = 
\left(
 \left|+\frac{7}{2}\right\rangle, 
 \left|-\frac{1}{2}\right\rangle, 
 \left|-\frac{9}{2}\right\rangle, 
 \left|-\frac{7}{2}\right\rangle, 
 \left|+\frac{1}{2}\right\rangle, 
 \left|+\frac{9}{2}\right\rangle, 
 \left|-\frac{5}{2}\right\rangle, 
 \left|+\frac{3}{2}\right\rangle, 
 \left|+\frac{5}{2}\right\rangle, 
 \left|-\frac{3}{2}\right\rangle
\right) 
%\end{eqnarray}
%where, $U$ is 
%\begin{eqnarray}
% U &=& 
\nonumber\\
\times
 \begin{pmatrix}
  M_1 & 0 & M_2 R^{(2)}(\alpha) & 0 & 0 & 0 \\
  0 & M_1 & 0 & -M_2 R^{(2)}(\alpha) & 0 & 0 \\
  0 & 0 & 0 & 0 & 0 & M_3 M_2 R^{(2)}(\alpha) \\
  0 & 0 & 0 & 0 & -M_3 M_2 R^{(2)}(\alpha) & 0 \\
 \end{pmatrix},
\end{eqnarray}
where,
\begin{eqnarray}
 M_1 &=& 
 \frac{1}{2\sqrt{6}}
 \begin{pmatrix}
   1 \\
   \sqrt{14} \\
   3
 \end{pmatrix},
\quad
 M_2 = 
 \frac{1}{2\sqrt{30}} 
 \begin{pmatrix}
   \sqrt{3} & -4\sqrt{7} \\
   \sqrt{42} & 2\sqrt{2} \\
   -5\sqrt{3} & 0 \\ 
 \end{pmatrix},
\quad
 M_3 = 
 \frac{1}{4\sqrt{6}}
 \begin{pmatrix}
   -5\sqrt{2} & 2\sqrt{7} & -3\sqrt{2} \\
   \sqrt{42} & 2\sqrt{3} & -\sqrt{42} 
 \end{pmatrix}.
\end{eqnarray}

\subsubsection{$\tilde{J} = 11/2$}
%The transformation matrix from the $\tilde{J}=11/2$ pseudospin states 
\begin{eqnarray}
\left(
 \left|\Gamma_6, -\frac{1}{2} \right\rangle,
 \left|\Gamma_6, +\frac{1}{2} \right\rangle,
 \left|\Gamma_7, -\frac{1}{2} \right\rangle,
 \left|\Gamma_7, +\frac{1}{2} \right\rangle,
 \left|\Gamma_8^{(1)}, -\frac{1}{2} \right\rangle,
 \left|\Gamma_8^{(2)}, -\frac{1}{2} \right\rangle,
 \left|\Gamma_8^{(1)}, +\frac{1}{2} \right\rangle,
 \left|\Gamma_8^{(2)}, +\frac{1}{2} \right\rangle,
\right.
\nonumber\\
\left.
 \left|\Gamma_8^{(1)}, -\frac{3}{2} \right\rangle,
 \left|\Gamma_8^{(2)}, -\frac{3}{2} \right\rangle,
 \left|\Gamma_8^{(1)}, +\frac{3}{2} \right\rangle,
 \left|\Gamma_8^{(2)}, +\frac{3}{2} \right\rangle
\right)
\nonumber\\
=
\left(
 \left|+\frac{7}{2}\right\rangle, 
 \left|-\frac{1}{2}\right\rangle, 
 \left|-\frac{9}{2}\right\rangle, 
 \left|-\frac{7}{2}\right\rangle, 
 \left|+\frac{1}{2}\right\rangle, 
 \left|+\frac{9}{2}\right\rangle, 
 \left|-\frac{5}{2}\right\rangle, 
 \left|+\frac{3}{2}\right\rangle, 
 \left|+\frac{11}{2}\right\rangle, 
%\right.
%\nonumber\\
%\left.
 \left|+\frac{5}{2}\right\rangle, 
 \left|-\frac{3}{2}\right\rangle, 
 \left|-\frac{11}{2}\right\rangle
\right) 
%\end{eqnarray}
%where,
%\begin{eqnarray}
% U &=&
\nonumber\\
\times
 \begin{pmatrix}
  M_1 & 0  & 0 & 0 & M_3 R^{(2)}(\beta) & 0 & 0 & 0 \\
  0 & -M_1 & 0 & 0 & 0 & M_3 R^{(2)}(\beta) & 0 & 0 \\
  0 & 0    & M_2  & 0 & 0 & 0 & 0 & M_4 M_3 R^{(2)}(\beta) \\
  0 & 0    & 0 & -M_2 & 0 & 0 & M_4 M_3 R^{(2)}(\beta) & 0 \\
 \end{pmatrix},
\end{eqnarray}
%$M_1$ etc. are
where,
\begin{eqnarray}
 M_1 &=& 
 \frac{1}{4\sqrt{3}}
 \begin{pmatrix}
   \sqrt{35} \\
   -\sqrt{6} \\
   \sqrt{7} 
 \end{pmatrix},
\quad
 M_2 = 
 \frac{1}{4\sqrt{3}}
 \begin{pmatrix}
   \sqrt{11} \\
   \sqrt{22} \\
  -\sqrt{15} \\
 \end{pmatrix},
% \frac{1}{4\sqrt{123}} 
% \begin{pmatrix}
%   -15\sqrt{5} & -14\sqrt{2} \\
%    9\sqrt{10} & -16 \\
%    \sqrt{33}  & -2\sqrt{330} 
% \end{pmatrix}
\nonumber\\
 M_3 &=& 
 \frac{1}{4\sqrt{123}}
 \begin{pmatrix}
  7\sqrt{5} & 12\sqrt{2} \\
 -\sqrt{42} & 4\sqrt{105} \\
  -41       &  0 
 \end{pmatrix},
\quad
 M_4 =
 \frac{1}{16\sqrt{3}} 
 \begin{pmatrix}
  -7\sqrt{3} &  -\sqrt{70} &  5\sqrt{15} \\ 
    \sqrt{6} & -2\sqrt{35} & -3\sqrt{30} \\ 
  -\sqrt{55} & -\sqrt{462} & -\sqrt{11} 
 \end{pmatrix}.
\end{eqnarray}

\subsubsection{$\tilde{J} = 13/2$}
%The transformation matrix from the $\tilde{J}=13/2$ pseudospin states 
\begin{eqnarray}
\left(
 \left|\Gamma_6, -\frac{1}{2} \right\rangle,
 \left|\Gamma_6, +\frac{1}{2} \right\rangle,
 \left|\Gamma_7^{(1)}, -\frac{1}{2} \right\rangle,
 \left|\Gamma_7^{(2)}, -\frac{1}{2} \right\rangle,
 \left|\Gamma_7^{(1)}, +\frac{1}{2} \right\rangle,
 \left|\Gamma_7^{(2)}, +\frac{1}{2} \right\rangle,
 \left|\Gamma_8^{(1)}, -\frac{1}{2} \right\rangle,
 \left|\Gamma_8^{(2)}, -\frac{1}{2} \right\rangle,
\right.
\nonumber\\
\left.
 \left|\Gamma_8^{(1)}, +\frac{1}{2} \right\rangle,
 \left|\Gamma_8^{(2)}, +\frac{1}{2} \right\rangle,
 \left|\Gamma_8^{(1)}, -\frac{3}{2} \right\rangle,
 \left|\Gamma_8^{(2)}, -\frac{3}{2} \right\rangle,
 \left|\Gamma_8^{(1)}, +\frac{3}{2} \right\rangle,
 \left|\Gamma_8^{(2)}, +\frac{3}{2} \right\rangle
\right)
\nonumber\\
 = 
\left(
 \left|+\frac{7}{2}\right\rangle, 
 \left|-\frac{1}{2}\right\rangle, 
 \left|-\frac{9}{2}\right\rangle, 
 \left|-\frac{7}{2}\right\rangle, 
 \left|+\frac{1}{2}\right\rangle, 
 \left|+\frac{9}{2}\right\rangle, 
 \left|-\frac{13}{2}\right\rangle, 
 \left|-\frac{5}{2}\right\rangle, 
 \left|+\frac{3}{2}\right\rangle, 
 \left|+\frac{11}{2}\right\rangle, 
\right.
\nonumber\\
\left.
 \left|+\frac{13}{2}\right\rangle, 
 \left|+\frac{5}{2}\right\rangle, 
 \left|-\frac{3}{2}\right\rangle, 
 \left|-\frac{11}{2}\right\rangle
\right) 
%\end{eqnarray}
%$U$ is defined by 
%\begin{eqnarray}
% U &=& 
\nonumber\\
\times
 \begin{pmatrix}
  M_1 & 0 & 0 & 0 & M_3 R^{(2)}(\beta) & 0 & 0 & 0 \\ 
  0 & M_1 & 0 & 0 & 0 & -M_3 R^{(2)}(\beta) & 0 & 0 \\
  0 & 0 & M_2 R^{(2)}(\alpha) & 0 & 0 & 0 & 0 & -M_4 M_3 R^{(2)}(\beta) \\
  0 & 0 & 0 & M_2 R^{(2)}(\alpha) & 0 & 0 & M_4 M_3 R^{(2)}(\beta) & 0 \\
 \end{pmatrix},
\end{eqnarray}
where,
\begin{eqnarray}
 M_1 &=& 
 \frac{1}{4} 
 \begin{pmatrix}
   \sqrt{3} \\
  -\sqrt{2} \\
   \sqrt{11} 
 \end{pmatrix},
\quad
 M_2 = \frac{1}{4\sqrt{114}}
 \begin{pmatrix}
    4\sqrt{10} & -\sqrt{429} \\
  -2\sqrt{286} &  \sqrt{15}  \\
             0 & 19\sqrt{3}  \\
   2\sqrt{130} & 3\sqrt{33}  \\
 \end{pmatrix},
\nonumber\\
 M_3 &=& 
 \frac{1}{4\sqrt{5}}
 \begin{pmatrix}
   \sqrt{33} & 4\sqrt{2} \\
  -\sqrt{22} & 4\sqrt{3} \\
  -5 & 0 \\
 \end{pmatrix},
\quad
 M_4 = \frac{1}{16\sqrt{6}}
 \begin{pmatrix}
  -\sqrt{143} & -\sqrt{858} & -\sqrt{39} \\
   5\sqrt{5}  & -3\sqrt{30} & -\sqrt{165} \\
   -21 & -5\sqrt{6} & \sqrt{33} \\
   7\sqrt{11} & -\sqrt{66} & -9\sqrt{3} \\
 \end{pmatrix}.
\end{eqnarray}

\subsubsection{$\tilde{J} = 15/2$}
%The transformation matrix from the $\tilde{J}=15/2$ pseudospin states 
\begin{eqnarray}
\left(
 \left|\Gamma_6, -\frac{1}{2} \right\rangle,
 \left|\Gamma_6, +\frac{1}{2} \right\rangle,
 \left|\Gamma_7, -\frac{1}{2} \right\rangle,
 \left|\Gamma_7, +\frac{1}{2} \right\rangle,
 \left|\Gamma_8^{(1)}, -\frac{1}{2} \right\rangle,
 \left|\Gamma_8^{(2)}, -\frac{1}{2} \right\rangle,
 \left|\Gamma_8^{(3)}, -\frac{1}{2} \right\rangle,
 \left|\Gamma_8^{(1)}, +\frac{1}{2} \right\rangle,
\right.
\nonumber\\
\left.
 \left|\Gamma_8^{(2)}, +\frac{1}{2} \right\rangle,
 \left|\Gamma_8^{(3)}, +\frac{1}{2} \right\rangle,
 \left|\Gamma_8^{(1)}, -\frac{3}{2} \right\rangle,
 \left|\Gamma_8^{(2)}, -\frac{3}{2} \right\rangle,
 \left|\Gamma_8^{(3)}, -\frac{3}{2} \right\rangle,
 \left|\Gamma_8^{(1)}, +\frac{3}{2} \right\rangle,
 \left|\Gamma_8^{(2)}, +\frac{3}{2} \right\rangle,
 \left|\Gamma_8^{(3)}, +\frac{3}{2} \right\rangle
\right)
\nonumber\\
=
\left(
 \left|+\frac{15}{2}\right\rangle, 
 \left|+\frac{7}{2}\right\rangle, 
 \left|-\frac{1}{2}\right\rangle, 
 \left|-\frac{9}{2}\right\rangle, 
 \left|-\frac{15}{2}\right\rangle, 
 \left|-\frac{7}{2}\right\rangle, 
 \left|+\frac{1}{2}\right\rangle, 
 \left|+\frac{9}{2}\right\rangle, 
\right.
\nonumber\\
\left.
 \left|-\frac{13}{2}\right\rangle, 
 \left|-\frac{5}{2}\right\rangle, 
 \left|+\frac{3}{2}\right\rangle, 
 \left|+\frac{11}{2}\right\rangle, 
 \left|+\frac{13}{2}\right\rangle, 
 \left|+\frac{5}{2}\right\rangle, 
 \left|-\frac{3}{2}\right\rangle, 
 \left|-\frac{11}{2}\right\rangle
\right) 
%\end{eqnarray}
%where,
%\begin{eqnarray}
% U &=& 
\nonumber\\
\times
 \begin{pmatrix}
  M_1 & 0 & 0 & 0 & M_3 R^{(3)}(\Omega) & 0 & 0 & 0 \\
  0 & -M_1 & 0 & 0 & 0 & M_3 R^{(3)}(\Omega) & 0 & 0 \\
  0 & 0 & M_2 & 0 & 0 & 0 & 0 & M_4 M_3 R^{(3)}(\Omega) \\
  0 & 0 & 0 & -M_2 & 0 & 0 & M_4 M_3 R^{(3)}(\Omega) & 0 \\
 \end{pmatrix},
\label{Eq:J152}
\end{eqnarray}
where, 
\begin{eqnarray}
 M_1 &=& 
 -\frac{1}{8\sqrt{3}}
 \begin{pmatrix}
  \sqrt{65} \\
  \sqrt{21} \\
  \sqrt{99} \\
  \sqrt{7} 
 \end{pmatrix},
\quad
 M_2 =
 \frac{1}{8\sqrt{3}}
 \begin{pmatrix} 
  \sqrt{77} \\
  \sqrt{65} \\
 -\sqrt{39} \\
 -\sqrt{11} 
 \end{pmatrix},
\quad
 M_3 =
 \frac{1}{24\sqrt{19}}
 \begin{pmatrix}
  -2\sqrt{266} & -2\sqrt{1430} & -\sqrt{455} \\
  0            & 0             & 57\sqrt{3}  \\
  0            & 48\sqrt{2}    & -3\sqrt{77} \\
  2\sqrt{2470} & -2\sqrt{154}  & -7 \\            
%  -\sqrt{\frac{21}{86}} &  3\sqrt{\frac{143}{3182}} & -\frac{1}{8}\sqrt{\frac{91}{111}} \\
%   \sqrt{\frac{65}{86}} & 3\sqrt{\frac{231}{15910}} & -\frac{7}{8\sqrt{185}} \\
%  0                     &    -\sqrt{\frac{86}{185}} & -\frac{1}{8}\sqrt{\frac{231}{185}} \\
%  0                     & 0                         & -\frac{1}{8}\sqrt{\frac{185}{3}} \\
 \end{pmatrix},
\nonumber\\
 M_4 &=&
 \frac{1}{64\sqrt{3}}
 \begin{pmatrix} 
   \sqrt{15}   &  7\sqrt{91} & -\sqrt{429}  & -3\sqrt{273} \\
   \sqrt{3003} & -7\sqrt{55} & -3\sqrt{105} &  3\sqrt{165} \\
   \sqrt{5005} & -\sqrt{33}  & -15\sqrt{7}  & -17\sqrt{11} \\
   \sqrt{105}  & 17\sqrt{13} & -\sqrt{3003} &  11\sqrt{39} \\
 \end{pmatrix}.
\end{eqnarray}
%$M_3$ are chosen so that all columns are orthonormal to each other and orthogonal to $M_1$.
%Thus, $M_3$ can be replaced by $M_3.V$ with arbitrary three-dimensional unitary matrix $V$.

\end{widetext}

\section{Decomposition of operator}
\label{A:ITO}
An operator $\hat{A}$ acting on the electronic states from $\mathcal{H}$ 
(\ref{Eq:H})
is decomposed into the irreducible tensor operators (\ref{Eq:Ykq}):
\begin{eqnarray}
 \hat{A} &=& \sum_{kq} a_{kq} \mathcal{Y}_{kq}. %(\tilde{\bm{J}}).
%\frac{Y_k^q(\tilde{\bm{J}})}{Y_k^0(\tilde{J})}.
\label{Eq:A}
\end{eqnarray}
Here, $\tilde{\bm{J}}$ is omitted for simplicity from the argument of $\mathcal{Y}_{kq}$, and the coefficients $a_{kq}$ are calculated as 
\begin{eqnarray}
 a_{kq} &=& (-1)^q 
 \frac{[k]}{[\tilde{J}]}
% \left(C_{\tilde{J}\tilde{J}k0}^{\tilde{J}\tilde{J}}\right)^2
 \left[\langle (\tilde{J}k) \tilde{J}\tilde{J}| \tilde{J}\tilde{J}k0\rangle \right]^2
% \text{Tr} \left[\frac{Y_{k}^{-q}(\tilde{\bm{J}})}{Y_k^0(\tilde{J})} \hat{A}\right],
 \text{Tr} \left[\mathcal{Y}_{k,-q} \hat{A}\right],
\label{Eq:akq}
\end{eqnarray}
$[x] = 2x + 1$, and $\text{Tr}$ is the trace over $\mathcal{H}$. %the pseudospin states. 
The irreducible tensor operators 
$\mathcal{Y}_{kq}$
are written in slightly different form compared to conventional Stevens operators \cite{Stevens1952}.
The advantages of the current form are that 
(a) explicit form of the Stevens operators is not necessary
(only easily obtainable Clebsch-Gordan coefficients are necessary), 
(b) it is suitable form for the use of group theoretical techniques, 
and (c) the coefficients $a_{kq}$ directly indicate the strength of the contribution because the magnitude of $\mathcal{Y}_{kq}$ is expected to be of the order of unity.

\begin{table}[tb]
\begin{ruledtabular}
\caption{Coefficients $v_{k,|q|}$ in Eq. (\ref{Eq:Vk}).}
\label{Table:V}
\begin{tabular}{cccccc}
$k$ & $v_{k,0}$ & $v_{k,4}$ & $v_{k,8}$ & $v_{k,12}$ & $v_{k,16}$ \\
\hline
 4 &  1 & $\sqrt{\frac{5}{14}}$ \\ 
 6 &  1 & $-\sqrt{\frac{7}{2}}$ \\
 8 &  1 & $\frac{1}{3}\sqrt{\frac{14}{11}}$ & $\frac{1}{3}\sqrt{\frac{65}{22}}$ \\
10 &  1 & $-\sqrt{\frac{66}{65}}$ & $-\sqrt{\frac{187}{130}}$ \\
12 &  1 & 0 & $\sqrt{\frac{429}{646}}$ & $4\sqrt{\frac{91}{7429}}$ \\
   &  0 & 1 & $-4\sqrt{\frac{42}{323}}$ & $9\sqrt{\frac{11}{7429}}$ \\
14 &  1 & $-\frac{3}{2}\sqrt{\frac{143}{595}}$ & $-\sqrt{\frac{741}{1190}}$ & $-\frac{1}{2}\sqrt{\frac{437}{119}}$ \\
16 &  1 & 0 & $\sqrt{\frac{442}{2185}}$ & $\frac{16}{5}\sqrt{\frac{17}{437}}$ & $7\sqrt{\frac{493}{135470}}$\\
   &  0 & 1 & $-6\sqrt{\frac{6}{805}}$ & $-\frac{31}{5}\sqrt{\frac{13}{483}}$ & $4\sqrt{\frac{754}{74865}}$ \\
\end{tabular}
\end{ruledtabular}
\end{table}

\section{The form of the crystal field}
\label{A:Hcf}
The totally symmetric $k$-th rank tensor of cubic group is expressed as 
\begin{eqnarray} 
 \hat{V}_k &=& v_{k,0} \mathcal{Y}_{k0}
 + \sum_{q=\pm 4} v_{k,4} \mathcal{Y}_{kq}
 + \sum_{q=\pm 8} v_{k,8} \mathcal{Y}_{kq}
\nonumber\\
 &+& \sum_{q=\pm 12} v_{k,12} \mathcal{Y}_{kq}
  +  \sum_{q=\pm 16} v_{k,16} \mathcal{Y}_{kq}.
% + \cdots. 
\label{Eq:Vk}
\end{eqnarray}
The coefficients $v_{k,|q|}$ listed in Table \ref{Table:V} are determined by making use of the fact that Eq. (\ref{Eq:Vk}) is invariant under $C_4^y$ and $C_4^z$ rotations.
The 12nd and 16th order operators contain two independent sets of coefficients which are shown in different lines in Table \ref{Table:V}.
The crystal field Hamiltonian is a linear combination of Eq. (\ref{Eq:Vk}):
\begin{eqnarray}
 \hat{H}_\text{cf} &=& \sum_k B_k \hat{V}_k. 
\end{eqnarray}
$B_0$ is the average of the crystal-field energies. 
There is one $B_k$ for each rank $k = 4, 6, 8, 10, 14$ and there are two $B_k$ for each $k = 12, 16$.

%\bibliography{ref}
%\end{document}


\end{document}